\documentclass[aps,pre,reprint,nofootinbib,groupedaddress]{revtex4-1}

\usepackage{appendix} 
\usepackage{amsmath}
\usepackage{amssymb}
\usepackage{graphicx}
\usepackage[ruled,vlined,noline]{algorithm2e}
\usepackage[ruled,vlined]{algorithm2e}
\begin{document}

\title{Evidential Reconstruction of Network from Time Series}



\author{Yishu Xian}
\affiliation{School of Science, Jiangsu University of Science and Technology}

\author{Zhaobo Zhang}
\affiliation{School of Science, Jiangsu University of Science and Technology}

\author{Cai Zhang}
\affiliation{School of Science, Jiangsu University of Science and Technology}


\author{Meizhu Li}
\email{Meizhu.li@ujs.edu.cn}
\affiliation{School of Computer Science and Communication Engineering, Jiangsu University}

\author{Qi Zhang}
\email{qi.zhang@just.edu.cn}
\affiliation{School of Science, Jiangsu University of Science and Technology}
\affiliation{Lorentz Institute for Theoretical Physics, Leiden University, PO Box 9504}


\date{\today}

\begin{abstract}
    Reconstructing the topology of complex networks from observational data remains a central challenge in network science. Here we propose a framework that is based on the Dempster-Shafer evidence theory to infer network structures directly from time series. By integrating multi-source information within an evidential reasoning scheme, the method captures underlying interaction patterns with high fidelity. Tests on three representative network models—Barabási-Albert Network, Erdős-Rényi Network, and Watts-Strogatz Network—show that the reconstruction accuracy is consistently high and remains robust against increases in network size and density. The application of the framework to real-world datasets from diverse domains further confirms its stability and general applicability. These results suggest that evidential reasoning offers a powerful and scalable approach for uncovering the structural organization of complex systems, especially when dealing with uncertain or incomplete multi-source data.
\end{abstract}


\maketitle

\section{Introduction}
Network science provides a methodological framework for characterizing the interactions among components within complex systems, with the primary goal of analyzing system behaviors and properties from a holistic perspective~\cite{newman2003structure}, as the focusing solely on individual units or pairwise independent interactions from a microscopic viewpoint often neglects the global association patterns and latent laws of the system~\cite{peel2022statistical}. According to its distinctive capabilities in uncovering dependencies between components, network theory has been extensively utilized in investigating complex systems within numerous disciplines, such as biological sciences~\cite{giaveri2024integrated,li2026protein}, social systems~\cite{zhang2026unveiling}, information dissemination~\cite{liu2026impact,wu2024influence}, transportation networks~\cite{ma2024distributional}, artificial intelligence~\cite{zhang2025leveraging,iacopini2026discovering}, and economic systems~\cite{fang2025utilizing}. Driven by the growing needs of research across diverse fields, network science has further extended into multiple subdomains of research, encompassing structural feature analysis of networks (such as community detection~\cite{jerdee2025normalized} and topological analysis~\cite{zhang2024structural}), network dynamics (including synchronization~\cite{scholes2024quantum}), robustness and resilience of networks~\cite{artime2024robustness} and higher-order network~\cite{xian2025topological,zhang2025uniform}, as well as theoretical explorations of network science grounded in statistical physics~\cite{giuffrida2025description,LI2026109369,zhang2023ensemble} and information theory within complex networks~\cite{somazzi2025learn}.

Among all of these research areas, accurately acquiring the topological structure of complex networks is a fundamental prerequisite for subsequent analysis and research. There are typically three different approaches to acquiring networks' structural information. One is based on network ensembles, the other one utilizing network dynamics and the time series that recods nodes' state change. Network ensemble methods aim to generate a set of networks that statistically resemble the observed network, thereby capturing its structural characteristics at a macro level~\cite{cimini2021reconstructing}. These methods often rely on specific constraints, such as degree sequences or strength sequences, to construct ensembles that reflect the underlying properties of the original network~\cite{squartini2017network,cimini2015systemic,cimini2021reconstructing}. The other one is reconstructing networks' dynamics that hidden in the network structure based on observed time series data. These methods leverage the dynamic behavior of nodes to uncover the underlying connections and interactions within the network. As for dynamics-based reconstruction methods, several main approaches have been developed, including those based on graph neural networks, Bayesian inference, compressive sensing, and model-driven reconstruction. Among these, reconstruction using graph neural networks typically requires obtaining the network structure and performing data training or achieves synchronous reconstruction by integrating prior knowledge with relationship identification~\cite{kipf2018neural,zhang2019general}. Similarly, Bayesian network reconstruction methods based on prior knowledge also rely on existing structural information to complete network inference~\cite{peixoto2018reconstructing,peixoto2019network}.
To achieve reconstruction under conditions where prior knowledge of the network structure is unavailable, researchers have gradually shifted toward extracting network topological information directly from time series. For instance, J. Casadiego et al. utilized linear regression equations for solving~\cite{casadiego2017model}, J. Runge et al. conducted causal network reconstruction based on time series~\cite{runge2018causal}, and T. P. Peixoto et al. employed subquadratic time algorithms to reduce complexity in network reconstruction~\cite{peixoto2025scalable}. The third approach focuses on the reconstruction of networks' topological structure from the time series of nodes' state changes. For instance, W. Wang et al., applied compressive sensing methods to more accurately extract network topological structure from time series for reconstruction~\cite{wang2016data}. Additionally, M. Chuang et al. performed network reconstruction from binary time series based on maximum likelihood estimation, further advancing reconstruction accuracy~\cite{ma2018statistical}. These reconstruction methods built a strong baseline for the application of network theory into the research on complex systems.

Network ensemble methods can reveal the structural characteristics of networks at a macro level, but they often struggle to provide precise topological information. To obtain more accurate network topology, dynamics-based reconstruction methods using observational data should be employed. Most of these methods rely on time series as the primary source of topological information. For example, it can reveal connections between users through likes or reposts on social platforms~\cite{eagle2009inferring}; constructing multi-omic network inference from time-series data~\cite{moscardo2025multi}; capture associations between political parties based on their votes for or against bills~\cite{zhang2008community}; infer the connectivity of neurons from changes in their activation states~\cite{yan2025efficient}; and reconstruct disease transmission pathways using epidemiological contact tracing data~\cite{pastor2015epidemic,bao2025epidemic}. In other words, the reconstruction of networks are the methods that deal with the uncertainty of connections between nodes.

However, those existing approaches are all design as an independent analytical frameworks, and there are typically build to deal with single-source information's network reconstruction. For instance, network ensemble methods often rely on specific constraints, such as degree sequences or strength sequences, to construct ensembles that reflect the underlying properties of the original network~\cite{squartini2017network,cimini2015systemic,cimini2021reconstructing}. Similarly, dynamics-based reconstruction methods using time series data typically focus on extracting topological information from a single source of time series. But in real-world scenarios, the information available for network reconstruction may multisources, for example, we may have degree and strength sequence simultaneously, we may also have several time series that record the states of nodes' infections that are start from different seeds nodes. In other words, the information used for network reconstruction should build on multi-source information, it fundamentally be regarded as a multi-source information fusion problem, rather than merely the continuous optimization of a single source of information.

As a mathematical framework intrinsically suited for handling uncertain information and inherently designed for information fusion, evidence theory provides a solid methodological foundation for related research~\cite{shafer2020mathematical,dempster1968upper,deng2015generalized}. Therefore, based on the fundamental principles of evidence theory, we propose an evidential reconstruction method to inference the topology of complex networks  from time series. The core idea of this method is to first extract multiple Basic Probability Assignments (BPA) from differents time series observations, which quantify the strength of association between system units. Then, by employing the Dempster combination rule, we integrate multiple time series BPAs to minimize the inherent uncertainty in the data. Finally, we reconstruct the network topology based on decision criteria derived from the fused evidence.
This method bulit an extensible framework for the topological structure reconstruction of complex networks, exhibiting favorable scalability and progressively improved performance: the longer the observed time series and the more diverse the data sources used for fusion, the more accurate the reconstructed network topology becomes. 

The structure of this paper is organized as follows: Section~\ref{se:METHODOLOGY} reviews the basic concepts and presents the overall framework and technical details of the proposed network's evidential reconstruction method; Section~\ref{se:perform} reports experimental results on synthetic networks (including Barabási-Albert(BA)~\cite{barabasi1999emergence}, Erdős-Rényi(ER)~\cite{erdds1959random}, Watts-Strogatz(WS) models)~\cite{watts1998collective} and real-world networks. Those reuslts provides the effective and accurate of our method. Section~\ref{se:result} discusses the validation of the reconstructed networks; and finally, Section~\ref{se:con} summarizes our main findings and concludes the paper.

\section{METHODOLOGY}\label{se:METHODOLOGY}
As we already mentioned in the introduction, the reconstruction of the networks' topological structure from time series can be fundamentally regarded as a multi-source uncertain information fusion problem. Therefore, based on the mathematical performance of the evidence theory in dealing with uncertain information and its inherent design for information fusion, our proposed method of the evidential reconstruction of networks is based on the evidence theory, which means our method does not require prior knowledge of network structure.
 
\begin{figure*}[htbp]
    \centering
    \includegraphics[width=\textwidth]{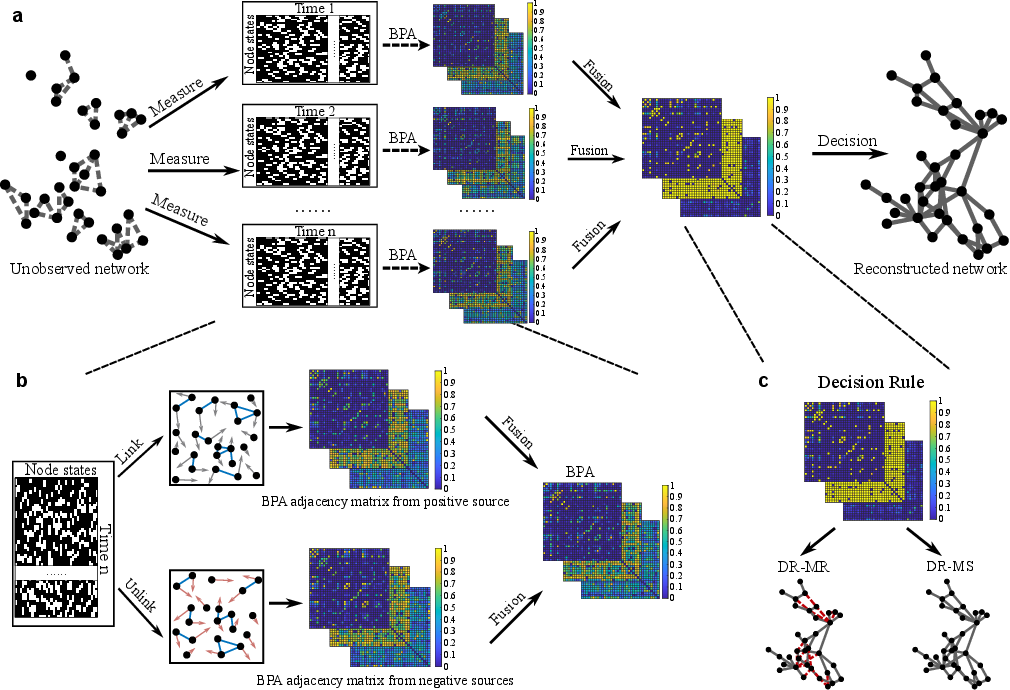}
    \caption{The process of evidential network reconstruction. The subfigure (a) shows the overall reconstruction procedure, in which Basic Probability Assignments (BPA) are constructed from time-series observations of an unknown network, followed by the fusion of multi-source time-series BPA, ultimately reconstructing the topology of the unknown network based on decision criteria. The subfigure (b) specifically demonstrates the extraction of two types of BPA—from both positive and negation perspectives-from a single time series, and their fusion into the BPA for that series. The subfigure (c) presents two decision methods: the Decision Rule Based on Minimum Robustness (DR-MR) and the Decision Rule Based on Maximum Similarity (DR-MS), with the former providing the upper bound of the decision range for the latter.}
    \label{jili}
\end{figure*}

Compared with existing methods that directly estimate edge probabilities from time series, the evidential reconstruction employs BPA to quantify the association strength between system units. The fusion rule based on evidence theory enables the integration of multiple time series, thereby minimizing the inherent uncertainty in the information. This design also leaves open the possibility for future reconstruction work that deals with data from diverse sources and perspectives (not limited to time series). Moreover, our method is not only straightforward to operate, but also ensures full transparency throughout the reconstruction process: every step of the fusion procedure, as well as the criteria and rationale for determining the final threshold, are explicitly traceable—constituting a "white-box" operation. The framework of the proposed methodology is illustrated in FIG.~\ref{jili}.

To facilitate the reader's understanding of the proposed network reconstruction framework, this section first reviews the basic principles of evidence theory and introduces relevant concepts related to binary time series, laying the groundwork for subsequent chapters. Then we detail the technical steps of the proposed evidential reconstruction method, including the extraction of BPA from time series, the fusion of multi-source BPAs, and the criteria for reconstructing network topology.

\subsection{Review of Basic Concepts}
Evidence theory, also known as Dempster-Shafer theory (DST), is a mathematical framework designed for handling uncertainty and ambiguous information, particularly well-suited for multi-source information fusion and decision-making analysis~\cite{shafer2020mathematical,Dempster1967Upper,deng2022random}. Its core principle involves defining a frame of discernment, where $\Theta$ denotes a finite universe of discourse (with its power set $2^{\Theta}$ forming the proposition space), and introducing a Basic Probability Assignment (BPA) function on this power set $2^{\Theta}$~\cite{deng2024RPSR}. The BPA assigns belief directly to sets rather than to individual elements, thereby enabling the expression of epistemic uncertainty. In multi-source information fusion, evidence theory combines pieces of evidence via the Dempster combination rule, which synthesizes evidence assignments by normalizing conflicts among them. The DST fundamental concepts are outlined below.
The frame of discernment is defined as $\Theta=\{\theta_1, \theta_2, ..., \theta_k\}$, a finite and complete set consisting of $k$ mutually exclusive elements. Its power set $2^{\Theta}$, comprising $2^k$ subsets, constitutes the set of all possible propositions.
\[
    2^\Theta = \{\varnothing,\{\theta_1\},\{\theta_2\}, ...,\{\theta_k\},\{\theta_1,\theta_2\}, ...,\{\theta_1,\theta_2,\theta_3\}, ...,\Theta\}.
\]
That is, the power set $2^{\Theta}$ is the collection of propositions. The Basic Probability Assignment (BPA) is then defined such that, for any subset $A \subseteq \Theta$, a function $m: 2^{\Theta} \to [0,1]$ satisfying the following two conditions:
\[
    m\{\varnothing\}= 0,
\]
and
\[
    \sum_{A \subseteq \Theta } m\{A\} = 1.
\]
is called a BPA or mass function. 

The value $m(A)$ is the basic probability assigned to proposition $A$ and represents the degree of belief allocated to $A$. Here, $A$ is any subset of the frame of discernment $\Theta$. If $m(A) > 0$, $A$ is referred to as a focal element of the BPA $m$ on $\Theta$.
The BPA is a central concept in DST, used to quantify the degree to which a piece of evidence supports different hypotheses. Specifically, BPA distributes the support or plausibility that each available evidence provides for a given hypothesis~\cite{deng2015generalized,li2024new}. Many methods exist for generating BPA, with the triangular fuzzy number approach being among the most commonly employed~\cite{zadeh1965fuzzy,liu2020determine}. Based on the BPA, we can construct a Belief function and a Plausibility function, which respectively provide the lower and upper bounds of confidence for a proposition.

The Dempster combination rule is the core fundamental rule in Dempster-Shafer theory. The combination rule for two independent pieces of evidence with BPAs $m_1$ and $m_2$ is expressed as follows:
\begin{equation}
    \left\{
    \begin{aligned}
    \quad m & \{\varnothing\}= 0, \\
    m &= \frac{1}{1-K}\sum_{A_i \cap B_i = A} m_1\{A_i\} \cdot m_2\{B_i\}, \\
    K &= \sum_{A_i \cap B_j = \varnothing} m_1\{A_i\} \cdot m_2\{B_j\}.
    \end{aligned}
    \right.
\end{equation}
where, $m_1$ and $m_2$ represent two groups of BPAs, and $K$ is referred to as the conflict coefficient (the larger the value of $K$, the greater the conflict; when $K = 1$, the combination rule becomes inapplicable).
This rule allows for the fusion of multiple evidence sources and computes the final degree of belief without requiring prior knowledge. Such an evidence synthesis method can effectively integrate information from multiple sources, particularly when the evidence is incomplete or ambiguous, offering distinct advantages over traditional probability theory~\cite{seiti2018risk,liu2019evidence}. Moreover, the fusion process is transparent and traceable.

In this study, the time series generated by binary-state dynamical models refers to observational sequences in which the state of each node in the time series matrix takes only two values: “inactive” (0) and “active” (1). Such binary time series can correspond to a variety of typical dynamical mechanisms, including evolutionary game models~\cite{szabo2007evolutionary}, threshold models~\cite{granovetter1978threshold}, and spreading (diffusion) models~\cite{kermack1927contribution,10.1214/aop/1176996493,zhan2025modeling}. Despite these binary dynamical processes generate from different model, but they share a common feature: the activation of a node is typically triggered by its interaction with already active nodes. This interaction relationship is precisely the key source of information for characterizing inter-node associations and inferring the underlying network structure. As the most widely used model in network science research, and are frequently employed both as benchmark tests for important node identification methods~\cite{chen2024identifying,yin2024identifying} and as tools for analyzing network dynamics~\cite{PhysRevE.61.5678,PhysRevE.111.024315}, this study adopts the Susceptible-Infected-Susceptible(SIS) model—a classical spreading model—as the method for generating time series data.

\subsection{Link information behind in Time Series}
As we mentioned before, the information contained in time series is the key to network reconstruction. Thus, the primary task is to acquire the time series data of the network, which serves as an indispensable foundation for the reconstruction process. The accuracy and completeness of the time series directly impact the reliability of subsequent reconstruction results. However, obtaining time series data is not the core focus of this study. Therefore, we opt for a simplified yet effective approach to simulate and generate the necessary foundational data. Specifically, we use the Susceptible-Infected-Susceptible (SIS) infectious disease propagation model to simulate the spread of a disease within a network, thereby swiftly obtaining corresponding contact tracing time series for the epidemic.
The method for generating time series data for the SIS infection model is as follows.

Step 1: Select a small number of nodes as the initial infection sources. These nodes will serve as the starting point for the disease spread within the network.

Step 2: For each infected node, identify its uninfected neighboring nodes. Using a predefined infection rate ($\beta$), each neighboring node has a certain probability of being infected. The disease is transmitted to these neighbors according to this probability.

Step 3: After each time step, apply the recovery process to the current set of infected nodes (denoted as $n_i$ ). With a specified recovery rate ($\gamma$), each infected node's state is changed from infected (I) to susceptible (S). It is important to note that the nodes newly infected in Step 2 are not included in the recovery process in Step 3.

Step 4: At the end of each time step, record the state of all nodes. This allows for tracking the evolution of the system over time.

Step 5: Repeat Steps 2 to 4 until one of the following conditions is met: the maximum number of infection time steps has been reached, all nodes have entered the infected state, or no new infection sources remain (i.e., no additional nodes can be infected in the subsequent steps).

This iterative process simulates the propagation of the infection throughout the network, and the recorded time series of node states can be used for network reconstruction and analysis. See the detailed description in Algorithm~\ref{Algorithm}.

\begin{algorithm}
\caption{Time Series Data Acquisition}
\label{Algorithm}
\textbf{Input:} Network $G=(V,E)$, infection rate $\beta$, recovery rate $\gamma$, maximum time step $T_{\max} = l$ \\
\textbf{Output:} Node state sequence $\mathbf{T_S} = \{\boldsymbol{t_s}(t)\}_{t=0}^{l} \in \{0,1\}^{l \times n}$

\SetKwFunction{Initialize}{Initialize}
\SetKwProg{Fn}{Function}{:}{}
\SetKw{KwTo}{to}
\SetKw{And}{and}

$t \leftarrow 0$\;
Select initial infected node set $V_{\text{inf}}(0) \subset V$\;
Initialize state vector: $\boldsymbol{t_s}(0,i) \gets \begin{cases} 1, & \text{if } i \in V_{\text{inf}}(0) \\ 0, & \text{otherwise} \end{cases}$\;

\While{$t < l$ \And $V_{\text{inf}}(t) \neq \emptyset$}{
  \For{$i \gets 1$ \KwTo $n$}{
    $\boldsymbol{t_s}(t+1,i) \gets \boldsymbol{t_s}(t,i)$\;
  }
  \ForEach{$i \in V_{\text{inf}}(t)$}{
    $\mathcal{N}_{\text{sus}}(i) \gets \{j \in \mathcal{N}(i) \mid \boldsymbol{t_s}(t,j)=0\}$\ \tcp*{$\mathcal{N}$ is the set of neighbors of infected nodes; $\mathcal{N}_{\text{sus}}(i)$ refers to the uninfected neighbors of node $i$.}
    \ForEach{$j \in \mathcal{N}_{\text{sus}}(i)$}{
      With probability $\beta$, set $\boldsymbol{t_s}(t+1,j) \gets 1$\;
    }
  }
  \ForEach{$i \in V_{\text{inf}}(t)$}{
    With probability $\gamma$, set $\boldsymbol{t_s}(t+1,i) \gets 0$\;
  }
  $V_{\text{inf}}(t+1) \gets \{i \in V \mid \boldsymbol{t_s}(t+1,i)=1\}$\;
  $t \leftarrow t + 1$\;
}
\end{algorithm}

Using the SIS model to acquire time-series data offers two key advantages. First, the propagation mechanism of the model is minimally constrained, which significantly enhances the applicability of the method. Second, if network reconstruction can be successfully achieved based on SIS-generated time-series data, it can be inferred that other types of time-series data—such as those derived from game-theoretic models or other propagation models—also hold reconstruction potential. Thus, as a flexible and efficient tool for simulating network time series, the SIS model ensures that the research findings can be effectively generalized to other models.

\subsection{Generation of the Mass Function}
The core step of the evidential reconstruction method is to extract the association relationships between nodes from time series and generate the corresponding BPA. 
We perform associative network inference on the time series of node states. Each pair of consecutive time steps forms an association unit, meaning a time series of length $l+1$ generates $l$ association units. 

\begin{figure}[htbp]
    \centering
    \includegraphics[width=8.6cm]{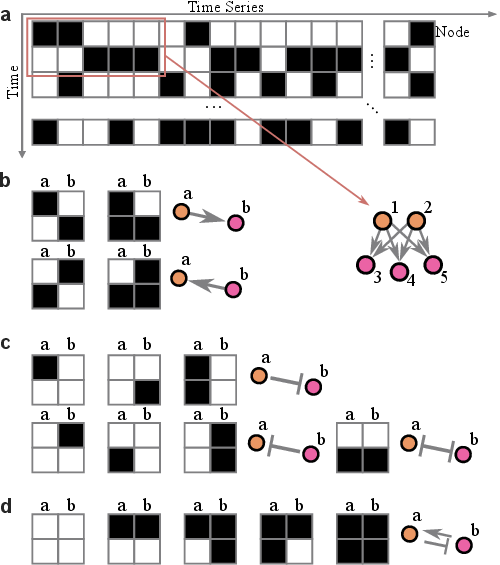}
    \caption{The association relationships between network nodes from time series. Specifically: (a) shows that nodes are considered associated in a given time step if they meet specific conditions; (b) outlines the criteria for determining association between nodes; (c) describes the conditions for confirming non-association; (d) represents cases of uncertainty, which provide no valid information regarding node associations.}
    \label{jz}
\end{figure}

Each association unit reveals the association between nodes, as illustrated in FIG.~\ref{jz}. For example, at time step t, nodes 1, and 2 are in the infected state (I); at time step t+1, nodes 3, 4, and 5 transition from susceptible (S) to infected (I). We consider nodes 3, 4, and 5 to have been infected by nodes 1, and 2. While nodes 3, 4, and 5 could have been infected by at most one node, we do not differentiate the specific sources. Instead, we treat them as being jointly infected by nodes 1, and 2.

\begin{figure}[htbp]
    \centering
    \includegraphics[width=8.6cm]{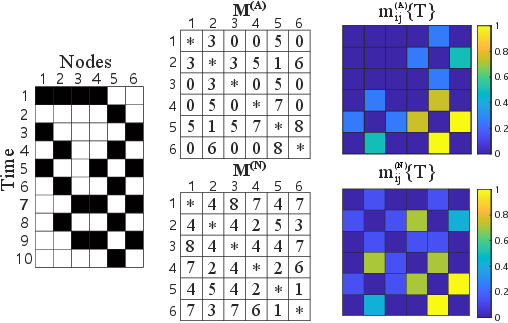}
    \caption{The process of generating $\mathbf{M^{(A)}}$ and $\mathbf{M^{(N)}}$ from time-eries data, and further constructing BPA matrices. Specifically, the $\mathbf{M^{(A)}}$ is built based on the number of associations between node pairs, while the $\mathbf{M^{(N)}}$ is derived from the number of non-associations between node pairs. Using these two matrices together with the BPA generation function (equations.~\ref{eq1}-\ref{eq6}), six BPA matrices are produced. Only two of them, $m^{(A)}\{T\}$ and $m^{(N)}\{T\}$, are shown in this figure.}
    \label{AAM}
\end{figure}

As shown in FIG.~\ref{AAM}, each time an infection occurs, we increment the value at the corresponding position in an $n\times n$ association adjacency matrix ($\mathbf{M^{(A)}}$) by one. This means that for $l$ association units, a maximum of $l$ times associations can be accumulated. At the same time, we construct a non-association adjacency matrix. If an association unit indicates no association between two nodes, we increment the value at the corresponding position in the non-association adjacency matrix ($\mathbf{M^{(N)}}$) by one.
In other words, let the time series be denoted as $\mathbf{T_S}\in\{0,1\}^{l\times n}$, and take the t-th row (the state at time t) as the row vector $\boldsymbol{t_s}(t)\in\{0,1\}^{1\times n}$.
For each time step t = 1, 2, ..., $l$-1, there are four state change vectors:
\begin{equation}
    \left\{
    \begin{aligned}
    \quad \boldsymbol{u}_{00}(t) & = (1-\boldsymbol{t_s}(t))\odot (1-\boldsymbol{t_s}(t+1)), \\
    \boldsymbol{u}_{01}(t) & = (1-\boldsymbol{t_s}(t))\odot \boldsymbol{t_s}(t+1), \\
    \boldsymbol{u}_{10}(t) & = \boldsymbol{t_s}(t)\odot (1-\boldsymbol{t_s}(t+1)), \\
    \boldsymbol{u}_{11}(t) & = \boldsymbol{t_s}(t)\odot \boldsymbol{t_s}(t+1).
    \end{aligned}
    \right.
\end{equation}
$\boldsymbol{u}_{00}$, $\boldsymbol{u}_{01}$, $\boldsymbol{u}_{10}$, and $\boldsymbol{u}_{11}$ denote the node infection state transitions as follows: 
    $\boldsymbol{u}_{00}$: transition from state $0$ to $0$ ($0\rightarrow0$), meaning the node remains uninfected;
    $\boldsymbol{u}_{01}$: transition from state $0$ to $1$ ($0\rightarrow1$), meaning the node becomes infected;
    $\boldsymbol{u}_{10}$: transition from state $1$ to $0$ ($1\rightarrow0$), meaning the node recovers to uninfected;
    $\boldsymbol{u}_{11}$: transition from state $1$ to $1$ ($1\rightarrow1$), meaning the node remains infected.
Thus, the matrix $\mathbf{M^{(A)}}$ is defined as
\begin{equation}
    \mathbf{M^{(A)}} = \sum_{t=1}^{l-1}{\Big[(\boldsymbol{a}(t))^\top \boldsymbol{u}_{01}(t)+\boldsymbol{u}_{01}(t)^\top (\boldsymbol{a}(t))\Big]}
\end{equation}
where \[\boldsymbol{a}(t) = \boldsymbol{u}_{10}(t)+\boldsymbol{u}_{11}(t)\]
And the matrix $\mathbf{M^{(N)}}$ can be defined as:
\begin{equation}
    \begin{split}
    \mathbf{M^{(N)}} = \sum_{t=1}^{l-1}\Big[
    &\boldsymbol{u}_{00}(t)^\top \boldsymbol{b}(t)
    + \boldsymbol{b}(t)^\top \boldsymbol{u}_{00}(t)\\
    &\quad + \boldsymbol{u}_{01}(t)^\top \boldsymbol{u}_{01}(t)
    \Big]
    \end{split}
    \end{equation}
where \[\boldsymbol{b}(t) = \boldsymbol{u}_{11}(t)+\boldsymbol{u}_{01}(t)+\boldsymbol{u}_{10}(t) = 1-\boldsymbol{u}_{00}(t)\]

We drew inspiration from Li et al.'s research on evidence identification for network influence to derive the method for generating BPA~\cite{li2018evidential}. 
Using the cumulative results from the $\mathbf{M^{(A)}}$ and $\mathbf{M^{(N)}}$, we obtain the basic probability assignments (denoted as  ${m^{(A)}}$ and  $m^{(N)}$, respectively) for each node pair. Since the network structure is entirely unknown, we employ a triangular fuzzy numbers method to generate the BPA.
For the association adjacency matrix, we assume that the node pair corresponding to the maximum value in the $\mathbf{M^{(A)}}$ has a 100\% confidence of being connected, while the pair corresponding to the minimum value among the off-diagonal elements has a 100\% confidence of being disconnected. This implies that the average of the maximum and minimum values corresponds to a single event (either $m^{(A)}\{T\}$ or $m^{(A)}\{F\}$) with a confidence of 0. We define this average value as the point where the confidence in the unknown state ($m^{(A)}\{T,F\}$) is 100\%, while at the maximum and minimum values themselves, the confidence in the unknown state is 0. 
Consequently, we derive the following three basic probability assignment functions for $\mathbf{M^{(A)}}$:
\begin{equation}
    m_{ij}^{(A)}\{T\} = \frac{M_{ij}^{(A)} - \mu_A}{\mathop{max}\limits_{(i,j)\in n}(M_{ij}^{(A)}) - \mu_A}
\label{eq1}
\end{equation}

\begin{equation}
    m_{ij}^{(A)}\{F\} = \frac{\mu_A - M_{ij}^{(A)}}{\mu_A - \mathop{min}\limits_{{(i,j)\in n}}(M_{ij}^{(A)})}
\label{eq2}
\end{equation}

\begin{equation}
    m_{ij}^{(A)}\{T, F\} = \frac{|\mu_A - M_{ij}^{(A)}\vert}{\Delta A / 2}
\label{eq3}
\end{equation}

Here, $\Delta_A$ represents the range of the matrix elements, given by \[\Delta A = \mathop{max}_{(i,j)\in n}M_{ij}^{(A)} - \mathop{min}_{(i,j)\in n}M_{ij}^{(A)},\] and the mean $\mu_A$ is defined as 
\[\mu_A = \frac{1}{2}{[\mathop{max}_{(i,j)\in n}(M_{ij}^{(A)}) +\mathop{min}_{(i,j)\in n}(M_{ij}^{(A)})]}.\] Since we do not consider self-loops, $i\neq j$

\begin{figure}[htbp]
    \centering
    \includegraphics[width=7cm]{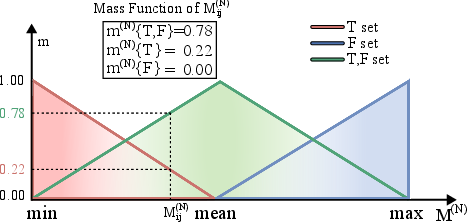}
    \caption{The generation of BPA from $\mathbf{M^{(N)}}$. It depicting how BPA is derived from a non-associated adjacency matrix. For a given node pair, if the cumulative association result is x, it corresponds to a belief assignment: $m^{(N)}\{T\}$  indicates that the degree of belief in the existence of an association between this node pair is 0.22, while $m^{(N)}\{T,F\}$ indicates that the degree of belief in being uncertain about its connection state is 0.78. Similarly, BPA from $\mathbf{M^{(A)}}$ can be generated in an analogous manner using this method.}
    \label{BPAmass}
\end{figure}

For the non-association adjacency matrix, which represents the adjacency of unconnected nodes, the maximum value corresponds to a 100\% confidence of being disconnected, while the minimum value corresponds to a 100\% confidence of being connected, as illustrated in FIG.~\ref{BPAmass}.

Thus, we obtain the following three basic probability assignment functions from $\mathbf{M_{N}}$:

\begin{equation}
    m_{ij}^{(N)}\{T\} = \frac{\mu_N - M_{ij}^{(N)}}{\mu_N - \mathop{min}\limits_{(i,j)\in n}(M_{ij}^{(N)})}
\label{eq4}
\end{equation}

\begin{equation}
    m_{ij}^{(N)}\{F\} = \frac{M_{ij}^{(N)} - \mu_N}{\mathop{max}\limits_{(i,j)\in n}(M_{ij}^{(N)}) - \mu_N}
\label{eq5}
\end{equation}

\begin{equation}
    m_{ij}^{(N)}\{T, F\} = \frac{|\mu_N - M_{ij}^{(N)}\vert}{\Delta N / 2}
\label{eq6}
\end{equation}
where, $\Delta_N$ represents the range of the matrix elements, given by \[\Delta N = \mathop{max}_{(i,j)\in n}(M_{ij}^{(N)}) - \mathop{min}_{(i,j)\in n}(M_{ij}^{(N)}),\] and the mean $\mu_N$ is defined as 
\[\mu_N = \frac{1}{2}{[\mathop{max}_{(i,j)\in n}(M_{ij}^{(N)}) + \mathop{min}_{(i,j)\in n}(M_{ij}^{(N)})]}.\]
In summary, we obtain a total of six basic probability assignment matrices through this process.

\subsection{Two levels of information fusion}
Regarding the fusion process of BPA, we actually perform fusion at two levels. The first step (Fusion 1) involves the internal fusion between $m^{(A)}$ and $m^{(N)}$, which are two observational means within the same time series. The second step (Fusion 2) involves the fusion process among BPAs obtained from multi-source time series. We will proceed to introduce the fusion methods for these two steps.

We use the most classic fusion rule from Dempster-Shafer theory to combine the aforementioned $m^{(A)}$ and $m^{(N)}$, that is~\cite{dempster1968upper}, 
\begin{equation}
    K = \sum_{X_i\bigcap Y_j=\varnothing} m^{(A)}\{X_i\} \cdot m^{(N)}\{Y_j\},
\end{equation}

\begin{equation}
    m(X) = \frac{1}{1-K}\sum_{X_i\bigcap Y_i=X} m^{(A)}\{X_i\} \cdot m^{(N)}\{Y_i\},
\end{equation}
Where $X_i$ and $Y_i$ represent the corresponding focal elements in $m^{(A)}$ and $m^{(N)}$, namely $m\{T\}$, $m\{F\}$, and $m\{T,F\}$. Where K is the conflict factor, which represents the degree of conflict between the two mass functions. Using these four equations, we obtain the fused mass function. The m\{T\} represents the confidence in the connection between a pair of nodes, and m\{F\}  represents the confidence in the disconnection between a pair of nodes. This fusion method combines the confidences derived from different perspectives of the same source's time series. However, this degree of fusion is not sufficient to meet the information requirements for network reconstruction. Therefore, we need to not only fuse the information obtained from different perspectives of a single time series but also fuse time series obtained from different infection sources. This is because time series obtained from different infection sources are significantly different. That is, different infection sources have different virus transmission paths, which lead us to obtain different time series.

Next, we need to fuse the mass functions from multi-source time series to obtain the final mass function used for decision-making. The process of multi-source fusion is key to reducing uncertain information. The associations between units of a complex system, derived from multiple time series, cross-validate one another, thereby pruning away uncertain connections to obtain a more accurate network topology. From the time series of k different infection sources, we can obtain their corresponding mass functions $m_1, m_2$, ...$m_{\mathrm{k}}$. Here, k represents the number of fused time series. We will fuse these mass functions to obtain the final mass function for decision-making. We fuse the obtained mass functions to derive the final mass function used for decision-making. The fusion rule is as follows:
\begin{equation}
    m = m_1\oplus  m_2\oplus  m_3 ...\oplus  m_\mathrm{k},
\end{equation}

\begin{figure}[htbp]
    \centering
    \includegraphics[width=8.6cm]{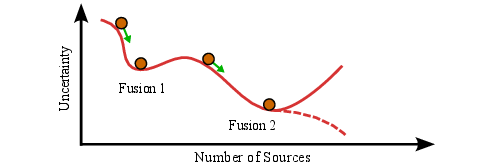}
    \caption{The changes in uncertainty during the fusion process of six BPA matrices. It demonstrates that the first fusion step (combining $m_A\{T\}$, $m_A\{F\}$, $m_A\{T,F\}$ with $m_N\{T\}$, $m_N\{F\}$, $m_N\{T,F\}$) reduces uncertainty. Subsequently, during the second fusion step (which integrates information from different time series), uncertainty may either increase or decrease—if contradictory time series are incorporated, uncertainty rises; otherwise, it continues to decrease.}
    \label{Fusion}
\end{figure}

Through two rounds of fusion across different dimensions, we obtained the final BPA decision matrix. Each cell in the matrix represents the connection belief between corresponding nodes. As shown in FIG.~\ref{Fusion}, the two-stage fusion process significantly reduces information uncertainty: first, the $\mathbf{M^{(N)}}$ and $\mathbf{M^{(A)}}$ information sources are fused, followed by a second fusion of BPAs generated from different infection sources, which further reduces overall uncertainty. Reducing information uncertainty is the core objective of our reconstruction method.

However, as shown in FIG.~\ref{Fusion}, increasing the number of information sources in the second fusion does not necessarily lead to a clearer understanding of the event. This is because multi-source information may contain both mutually corroborating and contradictory content. In fact, an increase in conflicting information can actually lead to higher uncertainty.

\subsection{The threshold probability in the decision rule}
Finally, we obtain the fused result, which is the belief value indicating whether each edge exists or not. Of course, after probability conversion, these edge existence beliefs can also be transformed into existence probabilities. However, within the traditional DST framework, event decisions can be made directly based on belief values. Typically, if the belief of an independent focal element exceeds 50\% or is the highest among all hypotheses, it can be regarded as the final result. This does not necessarily mean that this judgment method is optimal. As shown in subfigure (b) of FIG.~\ref{decision2}, such a decision approach may lead to a reconstructed network containing a large number of redundant edges.

We can define a minimum threshold based on network connectivity to determine that edges above this threshold exist, while edges below it do not. To more reasonably determine the belief decision threshold, we propose two threshold-based decision methods: one is based on minimum robustness, and the other is based on maximum similarity.

\subsubsection{Decision Rule Based on Minimum Robustness}
Decision Rule Based on Minimum Robustness(DR-MR) for obtaining the structural information of the network relies on the time series of SIS infections in the network. This means that we only classify nodes as part of the same network when they have all been infected by the same virus. Based on this fundamental idea, our approach aims to identify one or several edges with the minimum confidence that together connect the nodes of the network into a fully connected network, which we refer to as the minimum robust network. In this scenario, all edges with confidence values higher than these edges are considered to be present. The advantage of this method lies in its ability to prioritize the reconstruction of the edges with the highest confidence, while effectively discarding those with insufficient confidence, thereby significantly reducing the risk of reconstruction.

\begin{figure}[htbp]
    \centering
    \includegraphics[width=8.6cm]{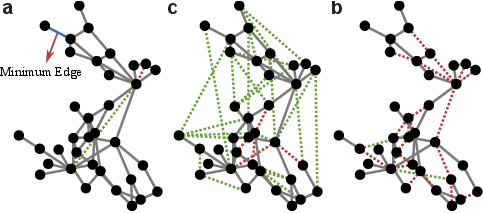}
    \caption{The process of the DR-MR decision method. Subfigure (a) shows the minimum edge confidence threshold to achieve network connectivity(edges above it are considered present, those below are absent). (b) indicates that too low a threshold introduces excessive redundant connections. (c) highlights a limitation: the DR-MR merely to ensure full connectivity may incorrectly classify many actual edges as absent.}
    \label{decision2}
\end{figure} 
In the associated adjacency matrix, the cumulative number of connections between pairs of nodes with a low degree or those that are farther from the hub nodes is usually lower than that of high-degree node pairs. As a result, the association between these nodes is weak, allowing them to serve as 'minimum points' in the BPA decision-making process. As shown in FIG.~\ref{decision2}, we can gradually lower the decision threshold until all nodes in the network can form a connected cluster. At this point, we can consider the network reconstruction to be complete, without the need to continue searching for additional 'minimum points.' In the experiment, we require the largest connected component to comprise at least 95\% of all nodes.

The core of the minimum-robustness approach lies in reconstructing only the high-confidence backbone, or "skeleton," of the network, rather than recovering its complete topology. Consequently, a primary limitation of this method is its tendency to produce an under-fitted network, as illustrated in subfigure (c) of FIG.~\ref{decision2}. Hence, this approach is more suitable for reconstructing networks that are inherently extremely sparse.

\subsubsection{Decision Rule Based on Maximum Similarity}
Although the minimum robustness decision method does not yield the most well-fitted network structure, it can help us determine an upper threshold. The Decision Rule Based on Maximum Similarity (DR-MS) is used to determine the optimal threshold for a time series by applying the Jaccard similarity coefficient. Before that, we need to review the Jaccard similarity coefficient.
\begin{equation}
    J = \frac{|E \cap \hat{E}|}{|E \cup \hat{E}|},
\end{equation}
When $J$ equals 1, it indicates that $E$ and $\hat{E}$ are completely identical. Conversely, when $J$ equals 0, it signifies that $T$ and $\hat{E}$ are completely dissimilar.

We employ a simple random sampling method to calculate the average Jaccard similarity. Specifically, we randomly select nodes that are infected at time step t+1 and exclude those already infected at time step t (i.e., removing non-recovered nodes), forming the set $T$. We then identify the set of neighbor nodes $\hat{T}$ in the reconstructed network. This process is repeated by randomly sampling $T$ at different time points, and the Jaccard similarity $J$ is computed each time. Different thresholds lead to different reconstructed networks, each corresponding to a different similarity value $J$.
Based on this, we progressively lower the decision threshold and fit the curve describing how $J$ varies with the threshold. The fitted $J$-curve exhibits a concave shape, and the peak of this curve corresponds to the threshold at which the reconstructed network most closely resembles the original time series. Using this optimal threshold, we can ultimately derive the network topology that best matches the original data.

To present the variation curve of the Jaccard coefficient ($J$) of the reconstructed network with respect to the threshold $\rho$ more clearly, we provide a rigorous mathematical proof of its concavity. The detailed proof is as follows:

\noindent\textbf{Notation:}

Let 
\[
a(\rho) = |E \cap \hat{E}(\rho)|, \quad b(\rho) = |E \cup \hat{E}(\rho)|,
\]
then \(J(\rho) = a(\rho)/b(\rho)\). Suppose \(\rho\) decreases from large to small values, and each time the threshold is lowered, the newly added edge set is \(\Delta\hat{E}\), which is disjoint from the existing \(\hat{E}(\rho)\). Define
\[
\Delta a = |E \cap \Delta\hat{E}|, \quad \Delta b = |\Delta\hat{E} \setminus E|.
\]
When the threshold decreases from \(\rho\) to \(\rho'\) (\(\rho' < \rho\)), we have
\[
a(\rho') = a(\rho) + \Delta a, \quad b(\rho') = b(\rho) + \Delta b.
\]

\noindent\textbf{Monotonicity Condition:}

Compare \(J(\rho')\) with \(J(\rho)\):
\[
J(\rho') = \frac{a(\rho) + \Delta a}{b(\rho) + \Delta b}.
\]
\(J(\rho') > J(\rho)\) if and only if
\[
\frac{a + \Delta a}{b + \Delta b} > \frac{a}{b} \iff b\Delta a > a\Delta b \iff \frac{\Delta a}{\Delta b} > \frac{a}{b} = J(\rho).
\]
Therefore, {\(J(\rho)\) increases if and only if \({\Delta a}/{\Delta b} > J(\rho)\), and decreases if and only if \({\Delta a}/{\Delta b} < J(\rho)\)}.

\noindent\textbf{Monotonicity of \({\Delta a}/{\Delta b}\):}

Consider the expansion process of \(\hat{E}(\rho)\) as \(\rho\) decreases. Since \(E\) is a fixed finite set, initially \(\hat{E}\) is empty. As it expands, the number of uncovered edges in \(E\) gradually decreases. Thus, in the early stage of expansion, the newly added edge set \(\Delta\hat{E}\) contains many edges from \(E\) (i.e., \(\Delta a\) is large), while as \(\hat{E}\) approaches the universal set, newly added edges mostly come from outside \(E\) (i.e., \(\Delta b\) increases and \(\Delta a\) decreases).

Define \(p = {\Delta a}/{|\Delta\hat{E}|}\), i.e., the proportion of newly added edges that belong to \(E\). Then
\[
\frac{\Delta a}{\Delta b} = \frac{p}{1-p}.
\]
Since \(p\) monotonically decreases as expansion proceeds (because \(E\) is finite and its uncovered portion decreases), \({p}/{1-p}\) also monotonically decreases. Therefore, \({\Delta a}/{\Delta b}\) monotonically decreases as \(\rho\) decreases (i.e., as expansion proceeds).

\noindent\textbf{Proof of Unimodality:}

\begin{enumerate}
    \item \textbf{Initial stage:} When \(\hat{E}\) is small, \({\Delta a}/{\Delta b}\) is large (typically \(\gg J(\rho)\), and \(J(\rho)\) is small), so \(J(\rho)\) increases.
    \item \textbf{Intermediate stage:} As expansion continues, \({\Delta a}/{\Delta b}\) gradually decreases. There exists some \(\rho^*\) such that
    \[
    \frac{\Delta a}{\Delta b} = J(\rho^*),
    \]
    at which point \(J(\rho)\) reaches its maximum.
    \item \textbf{Later stage:} When \(\rho < \rho^*\), \({\Delta a}/{\Delta b} < J(\rho)\), so \(J(\rho)\) decreases.
\end{enumerate}
Thus, \(J(\rho)\) attains a unique maximum at \(\rho^*\), and increases first then decreases, making it a unimodal function.

\noindent\textbf{Additional Remarks:}

In discrete cases, if at some step \(\Delta b = 0\) (i.e., all newly added edges belong to \(E\)), then \({\Delta a}/{\Delta b}\) can be regarded as infinite, still satisfying \({\Delta a}/{\Delta b} > J(\rho)\), so \(J\) increases. If \(\hat{E}\) already contains all of \(E\) (i.e., \(a\) no longer increases), then subsequently \(\Delta a = 0\), \({\Delta a}/{\Delta b} = 0 < J(\rho)\), and \(J\) decreases.

Under the assumption of continuous smoothness, unimodality implies that the function is quasi-concave, and under certain conditions it exhibits concavity, which ensures that the optimization problem is easy to solve.
In summary, \(J(\rho)\) exhibits unimodality with respect to \(\rho\), has a unique maximum, and the optimal threshold can be found using methods such as gradient ascent.

When the topological structure of the reconstructed network closely approximates the underlying true network, the maximum Jaccard similarity obtained (i.e., the peak value $J_{max}$ ) is numerically equal to the infection rate in the network propagation model. The process of parameter acquisition is as follows:

Let \(G=(V,E)\) be the true network, where \(V\) is the node set and \(E\) is the edge set. The SIS epidemic dynamics are governed by the infection rate \(\beta\) and recovery rate \(\gamma\).

Let \(V_{\text{inf}}(t) \subseteq V\) be the set of infected nodes at time \(t\), and \(V \setminus V_{\text{inf}}(t)\) the susceptible set. The expected number of nodes that remain infected from \(t\) to \(t+1\) is:
\[
\mathbb{E}\left[ |V_{\text{inf}}(t+1) \cap V_{\text{inf}}(t)| \,\big|\, V_{\text{inf}}(t) \right] = (1-\gamma)|V_{\text{inf}}(t)|.
\]
Define the observed recovery fraction at time \(t\) as:
\[
r_t = 1 - \frac{|V_{\text{inf}}(t+1) \cap V_{\text{inf}}(t)|}{|V_{\text{inf}}(t)|}.
\]
Then \(\mathbb{E}[r_t] = \gamma\). Averaging over \(M\) time steps yields a consistent estimator:
\[
\hat{\gamma} = \frac{1}{M}\sum_{t=1}^{M} r_t \overset{P}{\longrightarrow} \gamma \quad \text{as } M \to \infty.
\]

We reconstruct a network \(\hat{G}(\rho) = (V, \hat{E}(\rho))\) from the observed time series by tuning a threshold parameter \(\rho\). The Jaccard similarity between the true and reconstructed networks is:
\[
J(\rho) = \frac{|E \cap \hat{E}(\rho)|}{|E \cup \hat{E}(\rho)|}.
\]
Assume an optimal \(\rho^*\) maximizes \(J(\rho)\), and at this optimum \(\hat{E}(\rho^*) \approx E\). This assumption is justified a posteriori through validation.

Consider an infection event: let node \(v\) be infected at time \(t\). In the true network, the set of susceptible neighbors of \(v\) is \(\mathcal{N}(v) \cap (V \setminus V_{\text{inf}}(t))\). Each such neighbor becomes infected independently with probability \(\beta\). Let \(I_{\text{new}}(v,t)\) be the set of neighbors that actually become infected at \(t+1\). Then:
\[
\mathbb{E}\left[ |I_{\text{new}}(v,t)| \,\big|\, \mathcal{N}(v) \cap (V \setminus V_{\text{inf}}(t)) \right] = \beta \, |\mathcal{N}(v) \cap (V \setminus V_{\text{inf}}(t))|.
\]

In the reconstructed network, the susceptible neighbor set is \(\hat{\mathcal{N}}_\rho(v) \cap (V \setminus V_{\text{inf}}(t))\). Define the infection accuracy for this event as:
\[
J_{\text{inf}}(v,t;\rho) = \frac{|I_{\text{new}}(v,t)|}{|\hat{\mathcal{N}}_\rho(v) \cap (V \setminus V_{\text{inf}}(t))|}.
\]
Its conditional expectation is:
\[
\mathbb{E}\left[ J_{\text{inf}}(v,t;\rho) \,\big|\, \text{structure} \right] = \beta \, \frac{|\mathcal{N}(v) \cap (V \setminus V_{\text{inf}}(t))|}{|\hat{\mathcal{N}}_\rho(v) \cap (V \setminus V_{\text{inf}}(t))|}.
\]
If \(\hat{E}(\rho) \approx E\), then \(\hat{\mathcal{N}}_\rho(v) \approx \mathcal{N}(v)\), so:
\[
\mathbb{E}\left[ J_{\text{inf}}(v,t;\rho^*) \right] \approx \beta.
\]

Averaging over all infection events (\(k=1,\dots,K\)):
\[
\bar{J}_{\text{inf}}(\rho) = \frac{1}{M} \sum_{m=1}^{M} J_{\text{inf}}(v_k, t_k; \rho).
\]
By the law of large numbers, as \(M \to \infty\),
\[
\bar{J}_{\text{inf}}(\rho^*) \overset{P}{\longrightarrow} \beta.
\]

Thus, the infection rate can be estimated by maximizing \(\bar{J}_{\text{inf}}(\rho)\):
\[
\hat{\beta} = \max_\rho \bar{J}_{\text{inf}}(\rho).
\]
If the assumption is violated, meaning the reconstructed network significantly differs from the original network, then the obtained infection rate \(\beta\) will be incorrect. Consequently, when we later use this value to verify the credibility of the reconstructed network, we can easily detect that it is erroneous.

This conclusion establishes a direct bridge between network structural similarity and epidemiological dynamic parameters. After obtaining the infection and recovery rates, we can proceed to the final step of the framework: validating the credibility of the reconstructed network. Here, validation does not involve direct comparison with a known true network topology (since the true network is often unknown in real-world scenarios). Instead, it strictly relies on using only the time-series data to examine whether the reconstructed network and its inferred parameters can reproduce or predict the observed propagation dynamics. Specifically, we conduct propagation dynamics simulations based on the reconstructed network structure and the inferred parameters, and then compare the generated simulated time series with the original observed data, thereby assessing the reliability and predictive capability of the reconstruction results.
Of course, as the ultimate test of the method's validity, we will also conduct quantitative topological comparisons between the reconstructed and the true networks in benchmark cases where the ground-truth topology is known. This provides a comprehensive evaluation of the practical performance and advantages of the proposed framework.


\section{Reconstruction performance}\label{se:perform}
Before evaluating the performance of our method framework, we first need to determine how to quantify it. Network reconstruction can be performed at various scales, and what we define here is a reconstruction at the micro-scale. Therefore, we ultimately aim to obtain accurate associations between the units of the position network, i.e., an accurate network structure.
Accordingly, we define a method for assessing the performance of this approach—a criterion that can measure how similar the adjacency matrix of the reconstructed network ($\hat A$) is to the true underlying network behind the data. Similar to the positive indicator in Receiver Operating Characteristic (ROC) analysis, this criterion refers to the proportion of edges correctly identified in the reconstruction relative to the actual edges in the true network, namely:
\begin{equation}
    s = \frac{\sum_{i \neq j} \hat{a}_{ij} \cdot a_{ij} \cdot \delta(\hat{a}_{ij}) \cdot \delta(a_{ij})}{\sum_{i \neq j} a_{ij}}
\end{equation}
where $a_{ij}$ indicates whether nodes i and j are connected in the adjacency matrix of the original network, $\hat{a}_{ij}$ indicates whether nodes $i$ and $j$ are connected in the adjacency matrix of the reconstructed network.

Accordingly, we cannot rely solely on the positive metric; it is also necessary to consider the false positive metric, which refers to the number of edges reconstructed that do not exist in the true network. Considering that the adjacency matrices of real-world networks are almost always sparse, using the ratio of false positive edges to the number of non-existent edges in the true network is clearly unacceptable. This is because even if the number of extra edges is comparable to or several times greater than the number of true edges, this metric would still remain very small. Therefore, to more accurately characterize this metric, we define the following measure (redundancy rate):
\begin{equation}
    r = \frac{\sum_{i \neq j} \hat{a}_{ij} \cdot (1-a_{ij}) \cdot \delta(\hat{a}_{ij}) \cdot \delta(a_{ij})}{\sum_{i \neq j} a_{ij}}
\end{equation}
Using both the reconstruction rate and redundancy as metrics, we can assess the quantitative performance of our method. We will conduct reconstruction experiments on various network models with different structures, as well as on real-world networks, to evaluate the effectiveness of the approach.

\subsection{Evendentional Reconstruction of Karate-club network}
To help beginners quickly understand and become familiar with our methodological framework, we use a simple and classic network model for demonstration. The Karate Club network, collected by Zachary, describes the interactive relationships among 34 club members~\cite{zachary1977information}. First, we assume that we have no prior knowledge of the network's topological structure or other information. In fact, since its original interaction time series cannot be obtained, we simulate and generate the initial time series, which is treated as the most fundamental dataset. Using an infection rate ($\beta = 0.4$) and a recovery rate ($\gamma = 0.8$), we obtain its initial set of time series (k = 5). It should be noted that, in this example, the values of the infection and recovery rates are also considered unknown, reflecting a scenario where such prior information is unavailable.

We consider the time series to be reliable, and therefore, we do not address the issue of their reliability at this stage. We perform Basic Probability Assignment (BPA) on the statistical results of the time series for each edge according to equations.~\ref{eq1}-\ref{eq6}, and fill the resulting BPA values into the corresponding $n\times n$ association matrix set. Each time series generates six association matrices. The first fusion step involves combining $m^{(A)}$ and $m^{(N)}$.

\begin{figure*}[htbp]
    \centering
    \includegraphics[width=\textwidth]{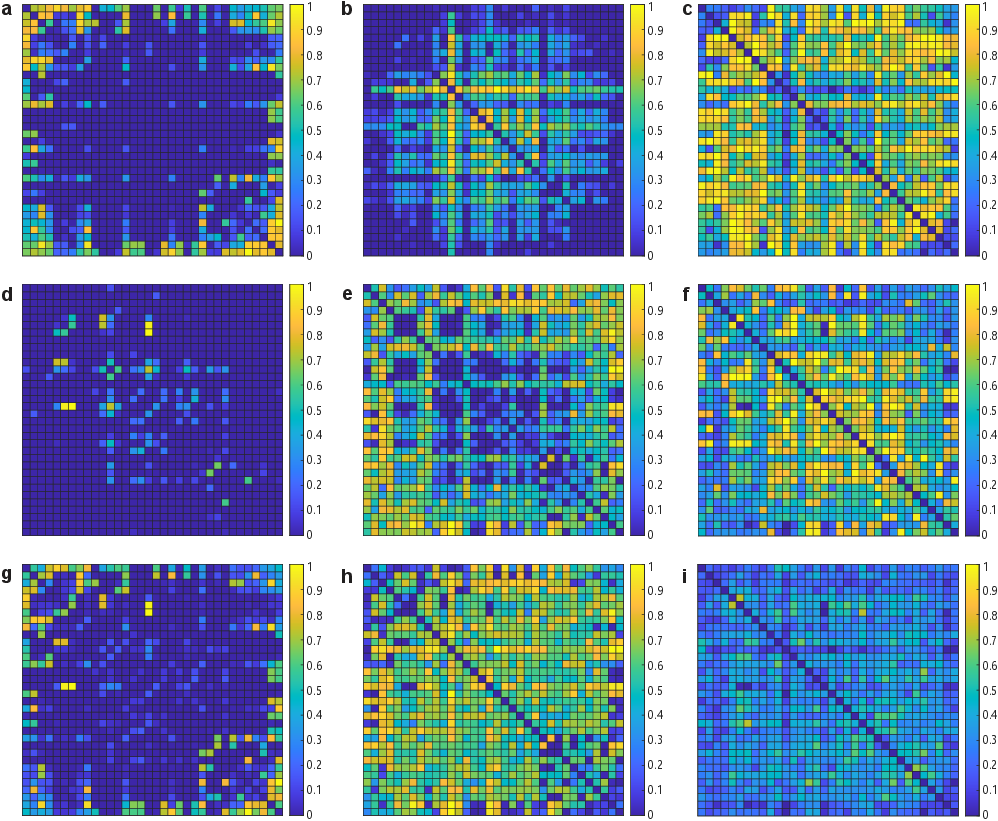}
    \caption{The heatmaps in subfigures (a-f) show the association matrices before and after the fusion of $\mathbf{M^{(A)}}$ and $\mathbf{M^{(N)}}$. Subfigures (a-c) display the heatmaps of the three association matrices m{T}, m{F}, and m{T, F} derived from the $\mathbf{M^{(A)}}$, while subfigures (d-f) correspond to the three heatmaps of m{T}, m{F}, and m{T, F} obtained from the $\mathbf{M^{(N)}}$. The last three subfigures present the fused association matrices.}
    \label{heatmap}
\end{figure*} 

In subfigures (a) and (d) of FIG.~\ref{heatmap}, the BPA matrices derived from $\mathbf{M^{(A)}}$ and $\mathbf{M^{(N)}}$ show that while it is difficult to directly infer node associations from a single association matrix, they can provide association information that is difficult to capture from $\mathbf{M^{(A)}}$ alone. Specifically, the edge with the highest belief in subfigure (d) is not detected in subfigure (a). It should be noted that this only represents the BPA matrix for m{T} obtained from two association matrices, while the other four matrices also carry important information. Thus, the role of $\mathbf{M^{(N)}}$ extends beyond this. As seen in subfigure (g), comparing the BPA matrix after the first fusion with the adjacency matrix of the original network reveals that $\mathbf{M^{(N)}}$ helps filter out certain ambiguous information and significantly reduces the belief in erroneous data. This demonstrates that the first fusion step aids in more accurately inferring the preliminary network structure.

\begin{figure}[htbp]
    \centering
    \includegraphics[width=8.6cm]{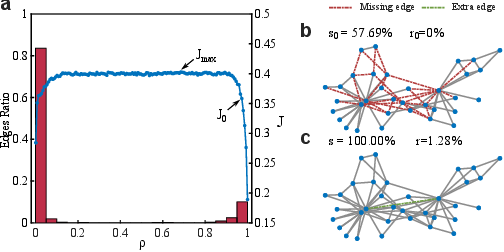}
    \caption{The network reconstruction results of two decision methods applied to the karate-club network. Among them, (a) shows the $J-\rho$ relationship curve (dotted line plot) derived from the Jaccard similarity calculation results, along with the corresponding histogram of edge distribution under the threshold $\rho$; (b) presents the network reconstructed based on the DR-MR method using the threshold $\rho_0$; and (c) displays the network structure reconstructed using the DR-MS method.}
    \label{Threshold}
\end{figure}

\begin{figure*}[htbp]
    \centering
    \includegraphics[width=17cm]{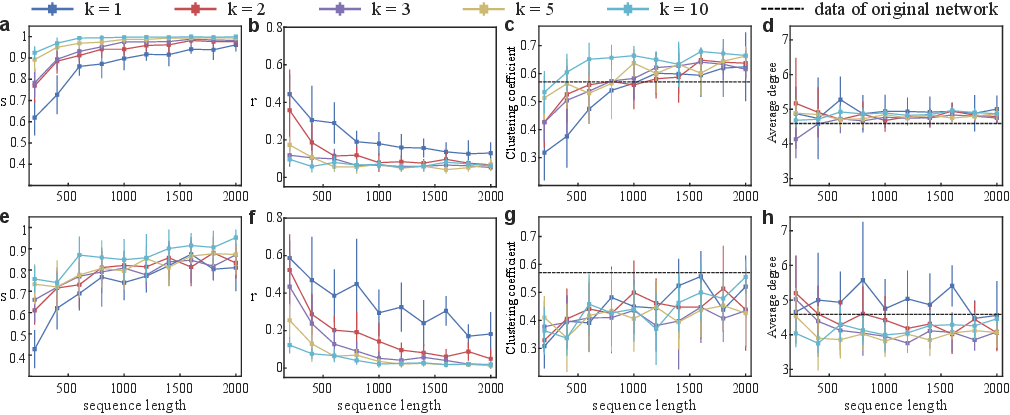}
    \caption{The accuracy of network reconstruction based on time series that generated under different infection parameters. Here, k represents the number of fused time series. (a-d) present the variation trends of four metrics with increasing series length under the infection parameters ($\beta_1=0.4$, $\gamma_1=0.7$), including the relationship between fusion frequency and reconstruction rate, the change in redundancy rate, as well as the evolution of average clustering coefficient and average degree. (e-h) display the corresponding metrics under the infection parameters ($\beta_2=0.2$, $\gamma_2=0.35$).}
    \label{figure_karate}
\end{figure*} 

The subsequent fusion process will be referred to as the second fusion. Specifically, this involves performing the second fusion on the first fused BPA matrix of each time series in the time series set. This fusion step is carried out by applying the traditional Dempster's combination rule, followed by normalization to obtain the final BPA matrix for decision-making. We prioritize the use of the minimum robustness decision method to identify the decision threshold $\rho _0$, at which the network just forms a relatively large connected graph. At this point, the network topology achieves $s_0$ = 57.69\% and $r_0$ = 0\%.
Thus, there remains one final step in our reconstruction process: determining the optimal threshold to reconstruct the network. Starting from $\rho _0$, we gradually decrease the threshold to obtain different network structures and compare the similarity of time series using $J$. We plot the $J-\rho$ curve and the distribution of edges under different thresholds, as shown in FIG.~\ref{Threshold}. By applying the $\rho$ value corresponding to $J_{max}$ (where $\rho  \leqslant  \rho _0$), the resulting network topology represents the final desired structure. The reconstruction achieves a satisfactory outcome, with a reconstruction rate s = 100.00\% and a redundancy rate r = 1.28\%.
The FIG.~\ref{Threshold} illustrates the distribution of edge confidence levels. Notably, the proportion of low-confidence edges is significant, exceeding 80\% of the total number of edges (i.e., approximately 449 out of the fully connected network's total edges M=561 have confidence levels approaching zero). As a result, over a relatively wide intermediate interval, the Jaccard similarity remains generally close to the maximum similarity $J_max$ =0.404, indicating that the network structures across this substantial middle range are largely similar. Leveraging the concavity of the similarity function, we can thereby determine the threshold that best matches the original time series and reconstruct the corresponding network topology.

It should be noted that for time series generated by the SIS infection model, different parameter combinations introduce varying degrees of uncertainty, and the length of the time series also affects the outcome of information extraction. To evaluate the applicability of our proposed method, we selected two sets of infection parameter configurations with the same basic reproduction number ($\beta_1=0.4$, $\gamma_1=0.7$ and $\beta_2=0.2$, $\gamma_2=0.35$) and compared their network reconstruction performance under varying lengths of time series.

However, even under infection parameter configurations with the same basic reproduction number, the information obtained is not entirely identical. For parameter sets characterized by high infection and high recovery rates, the resulting time series tend to reveal the network topology within a relatively short observation length. This is mainly because the high infection rate significantly reduces the required sequence length to traverse all possible network paths. Comparing the reconstruction performance under different infection parameters, networks reconstructed under high infection rates converge more quickly, but they also introduce more redundant information, leading to a relatively higher redundancy rate in the final reconstructed network. In contrast, time series with low infection rates can provide more accurate reconstruction information, but they require a longer sequence length to do so.

This observation is further supported by examining the topological structure of the reconstructed networks. Focusing on the optimal reconstruction results for the two infection parameter sets (sequence length = 2000, fusion count k = 10), the network reconstructed under high infection rates exhibits a higher clustering coefficient and average degree, meaning it contains a greater degree of redundancy. On the other hand, the network reconstructed based on low-infection-rate time series shows values closer to those of the original network in both metrics.

\subsection{Reconstruction on ER, BA and WS Networks}
A primary advantage of our method is that the network structural information is derived solely from the time series themselves, without relying on any prior knowledge of the underlying network structure, and the fusion process is well-defined. Therefore, our method can also be applied to help characterize complex systems in which direct observation of inter-unit relationships is difficult. However, is our methodological framework suitable for different topological structures? To validate the applicability of our framework, we constructed three classical network models with significantly distinct structures and used our method to reconstruct their networks. 

Three main network models are the Erdős-Rényi(ER) network, the Barabási-Albert(BA) network, and the Watts-Strogatz(WS) network. These models exhibit significant differences in their rules for node connections and network structures. The ER network is constructed based on random connections, where each time a new node is added, the probability of connecting to existing nodes is probability($P_{ER}$),
\[
    P_{\text{ER}}(i) = \frac{1}{N}
\]
with N representing the current number of nodes $i$ in the network. This means that each new node has an equal chance of connecting to any existing node, displaying strong randomness and uniformity. In contrast, the BA network has the characteristic of preferential attachment, where the probability of a new node connecting to an existing node is given by probability($P_{BA}$),
\[
    P_{\text{BA}}(i) = \frac{d(i)}{\sum d(i)}
\]
where $d(i)$ is the degree of the node $i$. This indicates that new nodes are more likely to connect to nodes with a higher degree, resulting in the formation of highly connected "hub" nodes, a property commonly observed in many real-world networks. The WS network establishes connections in a ring structure where each node is linked to a certain number of nearest neighbors (set to 4 in this experiment) and then randomly reconnects each edge with a fixed probability ($P_{WS}=constant$), maintaining one end while randomly selecting the other end to connect to any node in the network. This reconnection mechanism gives the WS network a small average path length and a high clustering coefficient, reflecting small-world characteristics, and it is widely used in the study of various phenomena in nature and society. The specific methods for generating these networks have been discussed in detail in the relevant literature~\cite{zhang2022betweenness,xian2025k}, so we will not elaborate further here. Since the infection parameters have a clear impact on the reconstruction results, we choose relatively appropriate parameters for the reconstruction.

\begin{figure}[htbp]
    \centering
    \includegraphics[width=8.6cm]{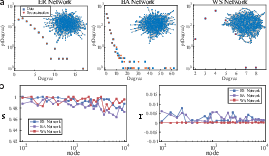}
    \caption{The reconstruction results of the network under BA, ER, and WS network models. (a) shows the degree distribution of the reconstructed network and the original network with 1000 nodes, as well as a visual representation of the network structure. (b) illustrates the relationship between network scale and both the reconstruction rate and redundancy rate for the three network models.}
    \label{model}
\end{figure} 
The subfigure (a) of FIG.~\ref{model} shows the visualized topological structures and their corresponding degree distribution curves for networks with 1,000 nodes. It can be observed that the degree distributions of the reconstructed networks align well with those of the original networks. Among them, the WS network demonstrates particularly remarkable reconstruction performance, as its degree distribution curve almost completely overlaps with that of the original network, further confirming the superior accuracy of this model in structural restoration. Although the reconstructed degree distributions of the ER and BA networks exhibit slight deviations in the tails, their overall distribution shapes remain well preserved.
Overall, the reconstruction rate is generally above 95\%, while the redundancy rate remains consistently below 1\%, indicating a high level of accuracy of the method. Notably, as shown in subfigure (b) of FIG.~\ref{model}, even when the network scale is expanded to 10,000 nodes, the reconstruction performance does not show a significant decline. This result further validates that the adopted method exhibits favorable scalability and stability.


Thus, we find that the reconstruction performance is not dependent on the specific type of topological structure. Specifically, whether applied to the BA scale-free network generated via preferential attachment, the entirely random ER network, or the highly clustered WS small-world network, the reconstruction results show consistently good performance with no significant differences observed. This indicates that our proposed method has broad applicability across networks with different topologies. Furthermore, experiments show that increasing the network size (i.e., the number of nodes) does not degrade the reconstruction performance, suggesting that the method maintains robustness even when applied to large-scale networks. Nevertheless, it remains to be examined whether reconstruction performance is influenced by network density, that is, the average degree $\langle  k\rangle$. To address this question, we will construct WS small-world network models with varying average degrees $\langle  k\rangle$ for further verification.
\begin{figure}[htbp]
    \centering
    \includegraphics[width=8.6cm]{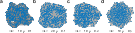}
    \caption{The structural visualization of a WS network under different combinations of average degree and rewiring probability.}
    \label{xing}
\end{figure} 
An important advantage of the WS network model generation rules is that they allow for convenient control of the average degree of the network, thereby providing a suitable basis for investigating the relationship between network density and reconstruction performance. The specific structure of the network is shown in FIG.~\ref{xing}.

\begin{figure}[htbp]
    \centering
    \includegraphics[width=8.6cm]{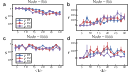}
    \caption{The relationship between network reconstruction performance and average degree. It presents the variation trends of reconstruction rate, redundancy rate, and the average degree of the original WS network for network sizes of 500 and 1000 nodes. In practice, the impact of network density on reconstruction effectiveness is minimal.}
    \label{dencity}
\end{figure}

The FIG.~\ref{dencity} illustrates the reconstruction performance for networks with N = 500 and N = 1000 nodes under different average degrees $\langle  k\rangle$, as well as under various rewiring probabilities in the WS network model. The results show that for the network with 500 nodes, as the average degree increases, the reconstruction rate slightly declines while the redundancy rate rises modestly. This trend aligns with expectations: given a fixed time-series length of 2000 and a fusion count of 5, the increase in the number of edges implies that more information is required to characterize the network topology, whereas the input information remains unchanged, leading to a mild degradation in reconstruction performance. In contrast, for the network with N = 1000, the decline in reconstruction performance with increasing average degree is more gradual. Moreover, different rewiring probabilities in the WS network do not exhibit a significant impact on the reconstruction results.

These findings provide a clear answer to the question raised earlier and further demonstrate that structural characteristics of the network(topological variations induced by changes in rewiring probability)do not substantially affect the reconstruction outcome.
Simulation results demonstrate that even as network density increases, the proposed reconstruction method continues to perform well, particularly when dealing with complex networks that exhibit local tightness and global connectivity, without significant degradation in stability or robustness.

\subsection{Reconstruction of Real-World Networks}

\begin{figure*}[htbp]
    \centering
    \includegraphics[width=\textwidth]{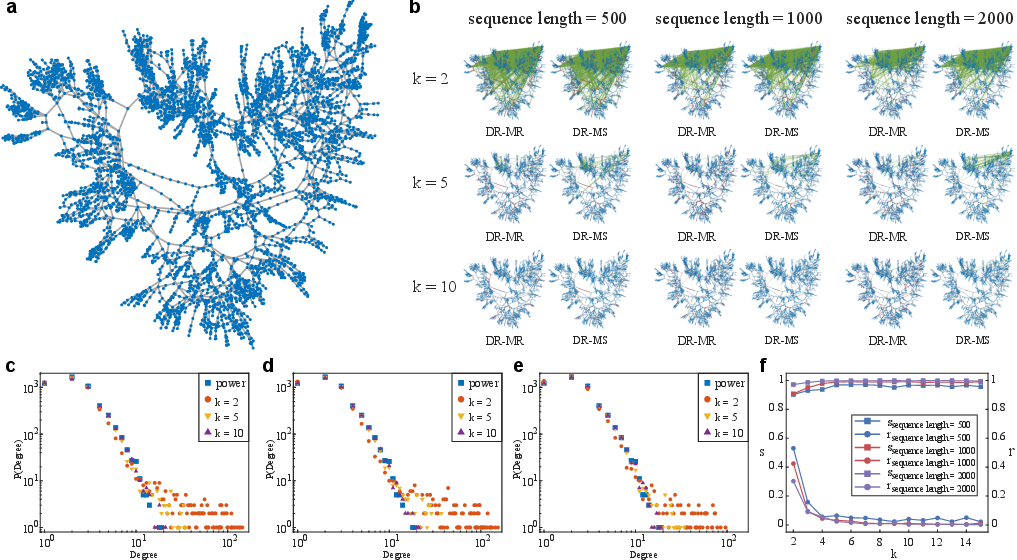}
    \caption{The reconstruction performance of the U.S. power network. (a) visualizes the topological structure of the original network; (b) presents the variations in reconstruction performance for two decision methods with increasing sequence length and fusion count, respectively; (c) to (e) show the degree distribution results of the reconstructed networks under sequence lengths of 500, 1000, and 2000; (f) compares the reconstruction performance under three sequence lengths at a fixed fusion count of k=10.}
    \label{tendency}
\end{figure*}

Building on the reconstruction performance demonstrated on simple networks and the three classical topological models, we further apply our method to real-world networks to validate its effectiveness in practical scenarios. To this end, we select several real networks~\cite{nr,cho2014wormnet,ahmed2010time,rossi2012fastclique,rossi2014pmc-www,truthy} from diverse domains—including social Networks, Transportation Networks, Power Networks, Gene Networks, and Biological Network—generate corresponding time series using the SIS infection model, and perform network reconstruction. Table.~\ref{tab:results} summarizes the structural characteristics, reconstruction parameters, and corresponding reconstruction performance of these networks. Overall, satisfactory reconstruction results are achieved across all real networks considered. Taking the U.S. Power network as an example, its specific reconstruction performance under two threshold decision methods is demonstrated in FIG.~\ref{tendency}. We preprocessed all the networks by retaining the largest connected subnetwork and removing all self-loops.

The U.S. Power Network consists of 4941 nodes and 6594 edges. It is relatively large in scale and exhibits topological characteristics similar to those of a BA scale-free network—its degree distribution shows significant heterogeneity, with a small number of hub nodes carrying a large portion of the connections. Given that many real-world infrastructure and social networks share similar scale-free properties, this case is particularly meaningful for validating the practical applicability of the proposed method.
As shown in subfigure (b) of FIG.~\ref{tendency}, during the early stage of network reconstruction (with only 2 fusion iterations), the reconstructed result still contains a high proportion of redundant edges, indicating that the structural information has not yet fully converged. Even when the time series length is increased to 2000, although the number of redundant edges decreases to some extent, the level of redundancy remains relatively high. 
This reflects that when information is insufficient or fusion is incomplete, the estimated network topology still includes considerable noise.
As the number of fusion iterations is further increased to 5, structural information continues to be integrated and refined, leading to a significant reduction in redundant edges. The network topology gradually converges to a stable and reasonable state. At this stage, the reconstructed network essentially meets the expected requirements in terms of connection accuracy, sparsity, and consistency with the original topology. This process demonstrates that when dealing with large-scale real-world networks with scale-free characteristics, appropriately increasing the number of fusion iterations can effectively enhance the reliability and structural clarity of the reconstruction results, further highlighting the robustness and adaptability of the method in practical scenarios. 

Readers may wonder whether increasing the number of fusion iterations essentially equates to extending the length of the time series, thereby naturally improving reconstruction performance. It should be noted that while additional fusion iterations inevitably introduce more temporal information, their impact on the reconstruction outcome is primarily attributed to the iterative refinement and noise suppression of structural information through multiple rounds of fusion, rather than merely accumulating more data. As can be observed in both subfigures (b) and (f), increasing the number of fusion iterations contributes far more significantly to reconstruction performance than simply extending the length of the time series.
Furthermore, Table.~\ref{tab:results} presents a comparative overview of the reconstruction results obtained by the two decision methods. The Minimum Robustness decision method (DR-MR) effectively suppresses the generation of edges that do not originally exist in the network, but it has a clear drawback: it leads to the loss of a large number of actual edges, resulting in insufficient reconstruction completeness. Taking the Metabolic network as an example, although the network reconstructed using the DR-MR method achieves a very low redundancy rate, its reconstruction rate is only 40.05\%, which is often unacceptable in practical applications. However, this method can identify the subset of edges with the highest confidence in the network, helping to clearly recognize node pairs with stronger associations and thereby revealing the most robust connections within the network.

\begin{table*}[htbp]
    \caption{\label{tab:results}
    The result of reconstruction for the real networks.}
    \begin{ruledtabular}
    \begin{tabular}{lccccccccc}
    &Nodes&Edges& k &$\langle k \rangle$&$\langle C \rangle$&$s_0$&$r_0$&s&r\\
    \colrule
    Karate-Club&34 &78&5 &4.5882 &0.5706 &52.56\% &0.00\% &100.00\% &0.00\%\\
    Celegans-Metabolic &453 &2025&9 &8.9404 &0.6465 &40.05\% &0.15\% &92.44\% & 3.56\% \\
    Twitter-Copen &761 &1029&5 &2.7043 &0.0098 &80.27\% &0.01\% &98.15\% &0.78\%\\
    C. elegans-Genes  &2194 &2688&6 &2.4503 &0.0098 &96.13\% &0.00\% &100.00\% &0.00\%\\
    US-Power &4941 &6594&8 &2.6691 &0.0801 &89.63\% &0.15\% &99.65\% &0.23\%\\
    Bahrain-Retweet&4676 &7979&9 &3.4127 &0.0172 &87.58\% &0.11\% &95.83\% &0.24\%\\
    Germany-Highway&1168 &1243&5 &2.1284 &0.0012 &95.49\% &0.00\% &99.92\% &0.64\%\\
    \end{tabular}
    \end{ruledtabular}
\end{table*}

In contrast, the Maximum Similarity decision method (DR-MS) achieves a better balance between reconstruction rate and redundancy rate. As shown in subfigures (c) to (e), even with a time series length of only 500, the degree distribution of the reconstructed network after 10 fusion iterations closely matches that of the original network.

\section{Result Validation}\label{se:result}
After completing the network reconstruction and comparing it with the original topological structure, the task is not yet finished. A critical point of concern is: in situations where the original network structure is unknown, on what basis can we trust that the reconstructed network truly represents the original one? To address this question, this paper proposes a comparative validation method based on relative entropy. This method involves counting the frequency of each node in the infected state from the original time series to derive the distribution of node infection states, and then uses relative entropy to measure the discrepancy between this distribution in the reconstructed network simulation and the original data. The core rationale is similar to methods used for validating node importance: hub nodes are typically infected first within a short time frame, which inherently reflects the intrinsic relationship between network structure and infection pathways (this point will not be elaborated further here).

Based on the reconstructed network topology and the inferred infection rates($\beta$) and recovery rates($\gamma$), we re-run propagation dynamics simulations to generate corresponding simulated time series. Subsequently, we calculate the relative entropy $D_{\text{KL}}$ between the node infection frequency distribution derived from the simulated series and that from the original series. To assess the acceptability of this value, we compare it with a baseline value, which is obtained by computing the relative entropy $D_{\text{KL}}$ between different repeated experiments on the original time series itself, reflecting the inherent statistical variability of the data. For single-source time series, we directly compute the relative entropy between the simulated time series and the original time series, and assess the trustworthiness of the reconstruction results based on the distribution difference in node infection frequencies.
\begin{equation}
    D_{\text{KL}}(P \parallel Q) = \sum_{i=1}^{N} P(i) \log \frac{P(i)}{Q(i)}
\end{equation}
where, $P(i)$ represents the distribution of infected states in the original time series, while $Q(i)$ denotes the distribution of infected states in the simulated time series generated from the reconstructed network. For multi-source time series, we compare the relative entropy of the infection frequency distributions within the original time series.
 
Specifically, for multi-source time series, we calculate the internal relative entropy by performing pairwise comparisons among all time series and computing their respective relative entropy. To minimize the impact of noise on the reliability of the reconstruction results as much as possible, we take the smallest relative entropy as the baseline value, denoted as $D_{KL}$. Subsequently, based on the reconstructed network and the estimated infection and recovery rates, we simulate and generate multiple sets of time series, and compute the relative entropy between each simulated series and the original time series. The smallest value among these is taken as $\hat D_{KL}$. The difference between the two relative entropy values is quantified using the following formula:
\begin{equation}
    \Delta D_{\text{KL}} = \frac {|D_{\text{KL}}-\hat{D}_{\text{KL}}\vert}{D_{\text{KL}}}.
\end{equation}
where $D_{KL}$ is the baseline relative entropy obtained from self-comparison of the original time series, and $\hat D_{KL}$ represents the relative entropy between the newly generated time series and the original time series.
The validation criterion is relatively straightforward: if the relative deviation between this relative entropy and the baseline relative entropy obtained from the original sequence itself is within 5\% ($\Delta D_{\text{KL}}\leqslant 5\%$), the reconstruction result is considered reliable.

\begin{figure}[htbp]
    \centering
    \includegraphics[width=8.6cm]{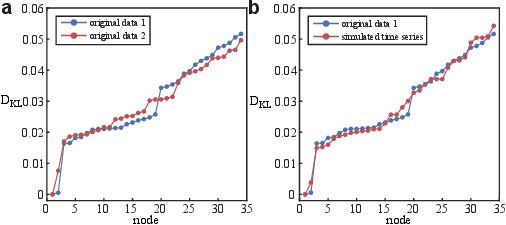}
    \caption{This figure compares the cumulative distribution of infection counts between the original network's time series and the simulated time series from the reconstructed network. The nodes are sorted according to the infection counts in the original network's time series. (a) shows the cumulative distribution of infection counts for different time series in the original network, while in (b), the blue dotted line represents the cumulative distribution corresponding to the original network's time series, and the red dotted line represents the cumulative distribution of the simulated infection series from the reconstructed network.}
    \label{KL}
\end{figure} 
As shown in FIG.~\ref{KL}, we conducted the calculation and validation of relative entropy for the time series based on the reconstructed karate-club network. Referring to FIG.~\ref{Threshold}, $J_{max}$ corresponds to the infection rate of the network. By substituting this parameter and applying the aforementioned method, the distribution of infection counts across nodes can be obtained. To clearly illustrate the distribution of node infection counts in FIG.~\ref{KL}, the 34 nodes are sorted in descending order of their infection counts. Based on this distribution, the relative entropy is calculated and compared with the relative entropy computed internally from the original time series. If the two values are very close, the reconstructed network can be considered credible. Under this relative entropy validation, if the frequency distribution of node infection states is highly consistent with the results obtained from the original time series($\Delta D_{\text{KL}}\leqslant 5\%$), the reconstruction result can be deemed correct. At this point, the network reconstruction task is finally considered complete.

\section{Conclusion}\label{se:con}
We propose a network reconstruction framework based on evidence theory, which does not rely on any prior structural information about the network. Instead, it achieves structural inference by extracting the implicit network topological features from time series data. Inferring node associations from time series data inherently involves significant uncertainty, and evidence theory can effectively handle such uncertainty, thereby improving the reliability of the reconstruction results. This method, which combines evidence theory with time series analysis to deduce network topology from data, offers new possibilities for exploring the structure of unknown complex systems. For complex systems where time series data is more readily available, this framework demonstrates good applicability and can reconstruct the network topology with considerable accuracy.

Experiments show that this method has a wide range of applications, strong stability, and is capable of reconstructing networks with various topological structures. We applied it to three classic network models with typical topologies and several real-world networks from different domains. In all test cases, it exhibited excellent reconstruction performance. Furthermore, we demonstrate the distinct roles of two-stage information fusion in reducing uncertainty and enhancing reconstruction effectiveness. Simultaneously, the study proves that the MR-MS method can also assist in estimating the infection parameters of the SIS epidemic model.

This framework is highly open and extensible, allowing for further expansion and generalization. For instance, the SIS time series data can be replaced with other information sources that contain network structures, such as time series derived from protein-protein interaction data, metabolic reaction pathways within organisms, or airline flight records, as well as time series obtained through multiple observation methods of the same network. The reconstruction target is not limited to static networks either—changes in network structure are reflected in time series data. By leveraging observational differences between time series from earlier and later networks, dynamic network reconstruction can be formulated. Even without prior knowledge of the network topology, this method can be adapted for reconstructing dynamic networks. Similarly, if partial local topological information or auxiliary information beyond the time series is available, even when it conflicts with inferences drawn from the time series, it can still be utilized through belief weighting or special processing to further enhance the accuracy and reliability of the reconstruction.

In summary, this framework provides a flexible and robust method for inferring network structure from time series data. By effectively integrating multi-source information and suppressing uncertainty through evidence theory, it shows considerable potential both theoretically and practically. It offers a new tool and perspective for the mining and dynamic analysis of complex system structures.
\section*{Acknowledgments}
This work is supported by the National Natural Science Foundation of China (Grant No. 62303198), The Research Initiation Fund for Senior Talents of Jiangsu University (No. 5501170008) and the Scientific Research Funding of Jiangsu University of Science and Technology (No.1052932204).

\bibliography{reconstruct.bib}

\begin{thebibliography}{72}%
\makeatletter
\providecommand \@ifxundefined [1]{%
 \@ifx{#1\undefined}
}%
\providecommand \@ifnum [1]{%
 \ifnum #1\expandafter \@firstoftwo
 \else \expandafter \@secondoftwo
 \fi
}%
\providecommand \@ifx [1]{%
 \ifx #1\expandafter \@firstoftwo
 \else \expandafter \@secondoftwo
 \fi
}%
\providecommand \natexlab [1]{#1}%
\providecommand \enquote  [1]{``#1''}%
\providecommand \bibnamefont  [1]{#1}%
\providecommand \bibfnamefont [1]{#1}%
\providecommand \citenamefont [1]{#1}%
\providecommand \href@noop [0]{\@secondoftwo}%
\providecommand \href [0]{\begingroup \@sanitize@url \@href}%
\providecommand \@href[1]{\@@startlink{#1}\@@href}%
\providecommand \@@href[1]{\endgroup#1\@@endlink}%
\providecommand \@sanitize@url [0]{\catcode `\\12\catcode `\$12\catcode `\&12\catcode `\#12\catcode `\^12\catcode `\_12\catcode `\%12\relax}%
\providecommand \@@startlink[1]{}%
\providecommand \@@endlink[0]{}%
\providecommand \url  [0]{\begingroup\@sanitize@url \@url }%
\providecommand \@url [1]{\endgroup\@href {#1}{\urlprefix }}%
\providecommand \urlprefix  [0]{URL }%
\providecommand \Eprint [0]{\href }%
\providecommand \doibase [0]{http://dx.doi.org/}%
\providecommand \selectlanguage [0]{\@gobble}%
\providecommand \bibinfo  [0]{\@secondoftwo}%
\providecommand \bibfield  [0]{\@secondoftwo}%
\providecommand \translation [1]{[#1]}%
\providecommand \BibitemOpen [0]{}%
\providecommand \bibitemStop [0]{}%
\providecommand \bibitemNoStop [0]{.\EOS\space}%
\providecommand \EOS [0]{\spacefactor3000\relax}%
\providecommand \BibitemShut  [1]{\csname bibitem#1\endcsname}%
\let\auto@bib@innerbib\@empty
\bibitem [{\citenamefont {Newman}(2003)}]{newman2003structure}%
  \BibitemOpen
  \bibfield  {author} {\bibinfo {author} {\bibfnamefont {M.~E.}\ \bibnamefont {Newman}},\ }\href@noop {} {\bibfield  {journal} {\bibinfo  {journal} {SIAM review}\ }\textbf {\bibinfo {volume} {45}},\ \bibinfo {pages} {167} (\bibinfo {year} {2003})}\BibitemShut {NoStop}%
\bibitem [{\citenamefont {Peel}\ \emph {et~al.}(2022)\citenamefont {Peel}, \citenamefont {Peixoto},\ and\ \citenamefont {De~Domenico}}]{peel2022statistical}%
  \BibitemOpen
  \bibfield  {author} {\bibinfo {author} {\bibfnamefont {L.}~\bibnamefont {Peel}}, \bibinfo {author} {\bibfnamefont {T.~P.}\ \bibnamefont {Peixoto}}, \ and\ \bibinfo {author} {\bibfnamefont {M.}~\bibnamefont {De~Domenico}},\ }\href@noop {} {\bibfield  {journal} {\bibinfo  {journal} {Nature Communications}\ }\textbf {\bibinfo {volume} {13}},\ \bibinfo {pages} {6794} (\bibinfo {year} {2022})}\BibitemShut {NoStop}%
\bibitem [{\citenamefont {Giaveri}\ \emph {et~al.}(2024)\citenamefont {Giaveri}, \citenamefont {Bohra}, \citenamefont {Diehl}, \citenamefont {Yang}, \citenamefont {Ballinger}, \citenamefont {Paczia}, \citenamefont {Glatter},\ and\ \citenamefont {Erb}}]{giaveri2024integrated}%
  \BibitemOpen
  \bibfield  {author} {\bibinfo {author} {\bibfnamefont {S.}~\bibnamefont {Giaveri}}, \bibinfo {author} {\bibfnamefont {N.}~\bibnamefont {Bohra}}, \bibinfo {author} {\bibfnamefont {C.}~\bibnamefont {Diehl}}, \bibinfo {author} {\bibfnamefont {H.~Y.}\ \bibnamefont {Yang}}, \bibinfo {author} {\bibfnamefont {M.}~\bibnamefont {Ballinger}}, \bibinfo {author} {\bibfnamefont {N.}~\bibnamefont {Paczia}}, \bibinfo {author} {\bibfnamefont {T.}~\bibnamefont {Glatter}}, \ and\ \bibinfo {author} {\bibfnamefont {T.~J.}\ \bibnamefont {Erb}},\ }\href@noop {} {\bibfield  {journal} {\bibinfo  {journal} {Science}\ }\textbf {\bibinfo {volume} {385}},\ \bibinfo {pages} {174} (\bibinfo {year} {2024})}\BibitemShut {NoStop}%
\bibitem [{\citenamefont {Li}\ \emph {et~al.}(2026)\citenamefont {Li}, \citenamefont {Li},\ and\ \citenamefont {Liu}}]{li2026protein}%
  \BibitemOpen
  \bibfield  {author} {\bibinfo {author} {\bibfnamefont {J.}~\bibnamefont {Li}}, \bibinfo {author} {\bibfnamefont {Y.~I.}\ \bibnamefont {Li}}, \ and\ \bibinfo {author} {\bibfnamefont {X.}~\bibnamefont {Liu}},\ }\href@noop {} {\bibfield  {journal} {\bibinfo  {journal} {Nature Genetics}\ ,\ \bibinfo {pages} {1}} (\bibinfo {year} {2026})}\BibitemShut {NoStop}%
\bibitem [{\citenamefont {Zhang}\ and\ \citenamefont {Xia}(2026)}]{zhang2026unveiling}%
  \BibitemOpen
  \bibfield  {author} {\bibinfo {author} {\bibfnamefont {M.}~\bibnamefont {Zhang}}\ and\ \bibinfo {author} {\bibfnamefont {Z.}~\bibnamefont {Xia}},\ }\href@noop {} {\bibfield  {journal} {\bibinfo  {journal} {Humanities and Social Sciences Communications}\ }\textbf {\bibinfo {volume} {13}},\ \bibinfo {pages} {67} (\bibinfo {year} {2026})}\BibitemShut {NoStop}%
\bibitem [{\citenamefont {Liu}\ \emph {et~al.}(2026)\citenamefont {Liu}, \citenamefont {Tang},\ and\ \citenamefont {Zhou}}]{liu2026impact}%
  \BibitemOpen
  \bibfield  {author} {\bibinfo {author} {\bibfnamefont {Y.}~\bibnamefont {Liu}}, \bibinfo {author} {\bibfnamefont {M.}~\bibnamefont {Tang}}, \ and\ \bibinfo {author} {\bibfnamefont {Y.}~\bibnamefont {Zhou}},\ }\href@noop {} {\bibfield  {journal} {\bibinfo  {journal} {Chaos, Solitons \& Fractals}\ }\textbf {\bibinfo {volume} {203}},\ \bibinfo {pages} {117715} (\bibinfo {year} {2026})}\BibitemShut {NoStop}%
\bibitem [{\citenamefont {Wu}\ and\ \citenamefont {Huo}(2024)}]{wu2024influence}%
  \BibitemOpen
  \bibfield  {author} {\bibinfo {author} {\bibfnamefont {B.}~\bibnamefont {Wu}}\ and\ \bibinfo {author} {\bibfnamefont {L.}~\bibnamefont {Huo}},\ }\href@noop {} {\bibfield  {journal} {\bibinfo  {journal} {Chaos, Solitons \& Fractals}\ }\textbf {\bibinfo {volume} {180}},\ \bibinfo {pages} {114522} (\bibinfo {year} {2024})}\BibitemShut {NoStop}%
\bibitem [{\citenamefont {Ma}\ and\ \citenamefont {Tang}(2024)}]{ma2024distributional}%
  \BibitemOpen
  \bibfield  {author} {\bibinfo {author} {\bibfnamefont {L.}~\bibnamefont {Ma}}\ and\ \bibinfo {author} {\bibfnamefont {Y.}~\bibnamefont {Tang}},\ }\href@noop {} {\bibfield  {journal} {\bibinfo  {journal} {Journal of International Economics}\ }\textbf {\bibinfo {volume} {148}},\ \bibinfo {pages} {103873} (\bibinfo {year} {2024})}\BibitemShut {NoStop}%
\bibitem [{\citenamefont {Zhang}\ \emph {et~al.}(2025{\natexlab{a}})\citenamefont {Zhang}, \citenamefont {Yang}, \citenamefont {Zenil}, \citenamefont {Chen}, \citenamefont {Shen}, \citenamefont {Kiani},\ and\ \citenamefont {Tegn{\'e}r}}]{zhang2025leveraging}%
  \BibitemOpen
  \bibfield  {author} {\bibinfo {author} {\bibfnamefont {H.}~\bibnamefont {Zhang}}, \bibinfo {author} {\bibfnamefont {C.-H.~H.}\ \bibnamefont {Yang}}, \bibinfo {author} {\bibfnamefont {H.}~\bibnamefont {Zenil}}, \bibinfo {author} {\bibfnamefont {P.-Y.}\ \bibnamefont {Chen}}, \bibinfo {author} {\bibfnamefont {Y.}~\bibnamefont {Shen}}, \bibinfo {author} {\bibfnamefont {N.~A.}\ \bibnamefont {Kiani}}, \ and\ \bibinfo {author} {\bibfnamefont {J.~N.}\ \bibnamefont {Tegn{\'e}r}},\ }\href@noop {} {\bibfield  {journal} {\bibinfo  {journal} {Nature Communications}\ } (\bibinfo {year} {2025}{\natexlab{a}})}\BibitemShut {NoStop}%
\bibitem [{\citenamefont {Iacopini}\ and\ \citenamefont {Valdano}(2026)}]{iacopini2026discovering}%
  \BibitemOpen
  \bibfield  {author} {\bibinfo {author} {\bibfnamefont {I.}~\bibnamefont {Iacopini}}\ and\ \bibinfo {author} {\bibfnamefont {E.}~\bibnamefont {Valdano}},\ }\href@noop {} {\bibfield  {journal} {\bibinfo  {journal} {Nature Computational Science}\ ,\ \bibinfo {pages} {1}} (\bibinfo {year} {2026})}\BibitemShut {NoStop}%
\bibitem [{\citenamefont {Fang}\ \emph {et~al.}(2025)\citenamefont {Fang}, \citenamefont {Chen}, \citenamefont {Zhou}, \citenamefont {Liu},\ and\ \citenamefont {Ge}}]{fang2025utilizing}%
  \BibitemOpen
  \bibfield  {author} {\bibinfo {author} {\bibfnamefont {C.}~\bibnamefont {Fang}}, \bibinfo {author} {\bibfnamefont {R.}~\bibnamefont {Chen}}, \bibinfo {author} {\bibfnamefont {L.}~\bibnamefont {Zhou}}, \bibinfo {author} {\bibfnamefont {X.}~\bibnamefont {Liu}}, \ and\ \bibinfo {author} {\bibfnamefont {S.}~\bibnamefont {Ge}},\ }\href@noop {} {\bibfield  {journal} {\bibinfo  {journal} {Habitat International}\ }\textbf {\bibinfo {volume} {161}},\ \bibinfo {pages} {103409} (\bibinfo {year} {2025})}\BibitemShut {NoStop}%
\bibitem [{\citenamefont {Jerdee}\ \emph {et~al.}(2025)\citenamefont {Jerdee}, \citenamefont {Kirkley},\ and\ \citenamefont {Newman}}]{jerdee2025normalized}%
  \BibitemOpen
  \bibfield  {author} {\bibinfo {author} {\bibfnamefont {M.}~\bibnamefont {Jerdee}}, \bibinfo {author} {\bibfnamefont {A.}~\bibnamefont {Kirkley}}, \ and\ \bibinfo {author} {\bibfnamefont {M.}~\bibnamefont {Newman}},\ }\href@noop {} {\bibfield  {journal} {\bibinfo  {journal} {Nature Communications}\ } (\bibinfo {year} {2025})}\BibitemShut {NoStop}%
\bibitem [{\citenamefont {Zhang}\ \emph {et~al.}(2024)\citenamefont {Zhang}, \citenamefont {Deng}, \citenamefont {Ding},\ and\ \citenamefont {Li}}]{zhang2024structural}%
  \BibitemOpen
  \bibfield  {author} {\bibinfo {author} {\bibfnamefont {Q.}~\bibnamefont {Zhang}}, \bibinfo {author} {\bibfnamefont {R.}~\bibnamefont {Deng}}, \bibinfo {author} {\bibfnamefont {K.}~\bibnamefont {Ding}}, \ and\ \bibinfo {author} {\bibfnamefont {M.}~\bibnamefont {Li}},\ }\href@noop {} {\bibfield  {journal} {\bibinfo  {journal} {Chaos, Solitons \& Fractals}\ }\textbf {\bibinfo {volume} {185}},\ \bibinfo {pages} {115158} (\bibinfo {year} {2024})}\BibitemShut {NoStop}%
\bibitem [{\citenamefont {Scholes}(2024)}]{scholes2024quantum}%
  \BibitemOpen
  \bibfield  {author} {\bibinfo {author} {\bibfnamefont {G.~D.}\ \bibnamefont {Scholes}},\ }\href@noop {} {\bibfield  {journal} {\bibinfo  {journal} {Proceedings of the Royal Society A}\ }\textbf {\bibinfo {volume} {480}},\ \bibinfo {pages} {20240209} (\bibinfo {year} {2024})}\BibitemShut {NoStop}%
\bibitem [{\citenamefont {Artime}\ \emph {et~al.}(2024)\citenamefont {Artime}, \citenamefont {Grassia}, \citenamefont {De~Domenico}, \citenamefont {Gleeson}, \citenamefont {Makse}, \citenamefont {Mangioni}, \citenamefont {Perc},\ and\ \citenamefont {Radicchi}}]{artime2024robustness}%
  \BibitemOpen
  \bibfield  {author} {\bibinfo {author} {\bibfnamefont {O.}~\bibnamefont {Artime}}, \bibinfo {author} {\bibfnamefont {M.}~\bibnamefont {Grassia}}, \bibinfo {author} {\bibfnamefont {M.}~\bibnamefont {De~Domenico}}, \bibinfo {author} {\bibfnamefont {J.~P.}\ \bibnamefont {Gleeson}}, \bibinfo {author} {\bibfnamefont {H.~A.}\ \bibnamefont {Makse}}, \bibinfo {author} {\bibfnamefont {G.}~\bibnamefont {Mangioni}}, \bibinfo {author} {\bibfnamefont {M.}~\bibnamefont {Perc}}, \ and\ \bibinfo {author} {\bibfnamefont {F.}~\bibnamefont {Radicchi}},\ }\href@noop {} {\bibfield  {journal} {\bibinfo  {journal} {Nature Reviews Physics}\ }\textbf {\bibinfo {volume} {6}},\ \bibinfo {pages} {114} (\bibinfo {year} {2024})}\BibitemShut {NoStop}%
\bibitem [{\citenamefont {Xian}\ \emph {et~al.}(2025{\natexlab{a}})\citenamefont {Xian}, \citenamefont {Chen}, \citenamefont {Li},\ and\ \citenamefont {Zhang}}]{xian2025topological}%
  \BibitemOpen
  \bibfield  {author} {\bibinfo {author} {\bibfnamefont {Y.}~\bibnamefont {Xian}}, \bibinfo {author} {\bibfnamefont {L.}~\bibnamefont {Chen}}, \bibinfo {author} {\bibfnamefont {M.}~\bibnamefont {Li}}, \ and\ \bibinfo {author} {\bibfnamefont {Q.}~\bibnamefont {Zhang}},\ }\href@noop {} {\bibfield  {journal} {\bibinfo  {journal} {Physics Letters A}\ ,\ \bibinfo {pages} {131233}} (\bibinfo {year} {2025}{\natexlab{a}})}\BibitemShut {NoStop}%
\bibitem [{\citenamefont {Zhang}\ \emph {et~al.}(2025{\natexlab{b}})\citenamefont {Zhang}, \citenamefont {Gao}, \citenamefont {Zhao}, \citenamefont {Hu}, \citenamefont {Miao},\ and\ \citenamefont {Zhang}}]{zhang2025uniform}%
  \BibitemOpen
  \bibfield  {author} {\bibinfo {author} {\bibfnamefont {K.}~\bibnamefont {Zhang}}, \bibinfo {author} {\bibfnamefont {J.}~\bibnamefont {Gao}}, \bibinfo {author} {\bibfnamefont {H.}~\bibnamefont {Zhao}}, \bibinfo {author} {\bibfnamefont {W.}~\bibnamefont {Hu}}, \bibinfo {author} {\bibfnamefont {M.}~\bibnamefont {Miao}}, \ and\ \bibinfo {author} {\bibfnamefont {Z.-K.}\ \bibnamefont {Zhang}},\ }\href@noop {} {\bibfield  {journal} {\bibinfo  {journal} {Physica A: Statistical Mechanics and its Applications}\ }\textbf {\bibinfo {volume} {666}},\ \bibinfo {pages} {130512} (\bibinfo {year} {2025}{\natexlab{b}})}\BibitemShut {NoStop}%
\bibitem [{\citenamefont {Giuffrida}\ \emph {et~al.}(2025)\citenamefont {Giuffrida}, \citenamefont {Squartini}, \citenamefont {Gr{\"u}nwald},\ and\ \citenamefont {Garlaschelli}}]{giuffrida2025description}%
  \BibitemOpen
  \bibfield  {author} {\bibinfo {author} {\bibfnamefont {F.}~\bibnamefont {Giuffrida}}, \bibinfo {author} {\bibfnamefont {T.}~\bibnamefont {Squartini}}, \bibinfo {author} {\bibfnamefont {P.}~\bibnamefont {Gr{\"u}nwald}}, \ and\ \bibinfo {author} {\bibfnamefont {D.}~\bibnamefont {Garlaschelli}},\ }\href@noop {} {\bibfield  {journal} {\bibinfo  {journal} {Physical Review Research}\ }\textbf {\bibinfo {volume} {7}},\ \bibinfo {pages} {043057} (\bibinfo {year} {2025})}\BibitemShut {NoStop}%
\bibitem [{\citenamefont {Li}\ and\ \citenamefont {Zhang}(2026)}]{LI2026109369}%
  \BibitemOpen
  \bibfield  {author} {\bibinfo {author} {\bibfnamefont {M.}~\bibnamefont {Li}}\ and\ \bibinfo {author} {\bibfnamefont {Q.}~\bibnamefont {Zhang}},\ }\href@noop {} {\bibfield  {journal} {\bibinfo  {journal} {Communications in Nonlinear Science and Numerical Simulation}\ }\textbf {\bibinfo {volume} {152}},\ \bibinfo {pages} {109369} (\bibinfo {year} {2026})}\BibitemShut {NoStop}%
\bibitem [{\citenamefont {Zhang}\ and\ \citenamefont {Garlaschelli}(2023)}]{zhang2023ensemble}%
  \BibitemOpen
  \bibfield  {author} {\bibinfo {author} {\bibfnamefont {Q.}~\bibnamefont {Zhang}}\ and\ \bibinfo {author} {\bibfnamefont {D.}~\bibnamefont {Garlaschelli}},\ }\href@noop {} {\bibfield  {journal} {\bibinfo  {journal} {Chaos, Solitons \& Fractals}\ }\textbf {\bibinfo {volume} {172}},\ \bibinfo {pages} {113546} (\bibinfo {year} {2023})}\BibitemShut {NoStop}%
\bibitem [{\citenamefont {Somazzi}\ and\ \citenamefont {Garlaschelli}(2025)}]{somazzi2025learn}%
  \BibitemOpen
  \bibfield  {author} {\bibinfo {author} {\bibfnamefont {A.}~\bibnamefont {Somazzi}}\ and\ \bibinfo {author} {\bibfnamefont {D.}~\bibnamefont {Garlaschelli}},\ }\href@noop {} {\bibfield  {journal} {\bibinfo  {journal} {Physical Review Research}\ }\textbf {\bibinfo {volume} {7}},\ \bibinfo {pages} {033087} (\bibinfo {year} {2025})}\BibitemShut {NoStop}%
\bibitem [{\citenamefont {Cimini}\ \emph {et~al.}(2021)\citenamefont {Cimini}, \citenamefont {Mastrandrea},\ and\ \citenamefont {Squartini}}]{cimini2021reconstructing}%
  \BibitemOpen
  \bibfield  {author} {\bibinfo {author} {\bibfnamefont {G.}~\bibnamefont {Cimini}}, \bibinfo {author} {\bibfnamefont {R.}~\bibnamefont {Mastrandrea}}, \ and\ \bibinfo {author} {\bibfnamefont {T.}~\bibnamefont {Squartini}},\ }\href@noop {} {\emph {\bibinfo {title} {Reconstructing networks}}}\ (\bibinfo  {publisher} {Cambridge University Press},\ \bibinfo {year} {2021})\BibitemShut {NoStop}%
\bibitem [{\citenamefont {Squartini}\ \emph {et~al.}(2017)\citenamefont {Squartini}, \citenamefont {Cimini}, \citenamefont {Gabrielli},\ and\ \citenamefont {Garlaschelli}}]{squartini2017network}%
  \BibitemOpen
  \bibfield  {author} {\bibinfo {author} {\bibfnamefont {T.}~\bibnamefont {Squartini}}, \bibinfo {author} {\bibfnamefont {G.}~\bibnamefont {Cimini}}, \bibinfo {author} {\bibfnamefont {A.}~\bibnamefont {Gabrielli}}, \ and\ \bibinfo {author} {\bibfnamefont {D.}~\bibnamefont {Garlaschelli}},\ }\href@noop {} {\bibfield  {journal} {\bibinfo  {journal} {Applied Network Science}\ }\textbf {\bibinfo {volume} {2}},\ \bibinfo {pages} {3} (\bibinfo {year} {2017})}\BibitemShut {NoStop}%
\bibitem [{\citenamefont {Cimini}\ \emph {et~al.}(2015)\citenamefont {Cimini}, \citenamefont {Squartini}, \citenamefont {Garlaschelli},\ and\ \citenamefont {Gabrielli}}]{cimini2015systemic}%
  \BibitemOpen
  \bibfield  {author} {\bibinfo {author} {\bibfnamefont {G.}~\bibnamefont {Cimini}}, \bibinfo {author} {\bibfnamefont {T.}~\bibnamefont {Squartini}}, \bibinfo {author} {\bibfnamefont {D.}~\bibnamefont {Garlaschelli}}, \ and\ \bibinfo {author} {\bibfnamefont {A.}~\bibnamefont {Gabrielli}},\ }\href@noop {} {\bibfield  {journal} {\bibinfo  {journal} {Scientific reports}\ }\textbf {\bibinfo {volume} {5}},\ \bibinfo {pages} {15758} (\bibinfo {year} {2015})}\BibitemShut {NoStop}%
\bibitem [{\citenamefont {Kipf}\ \emph {et~al.}(2018)\citenamefont {Kipf}, \citenamefont {Fetaya}, \citenamefont {Wang}, \citenamefont {Welling},\ and\ \citenamefont {Zemel}}]{kipf2018neural}%
  \BibitemOpen
  \bibfield  {author} {\bibinfo {author} {\bibfnamefont {T.}~\bibnamefont {Kipf}}, \bibinfo {author} {\bibfnamefont {E.}~\bibnamefont {Fetaya}}, \bibinfo {author} {\bibfnamefont {K.-C.}\ \bibnamefont {Wang}}, \bibinfo {author} {\bibfnamefont {M.}~\bibnamefont {Welling}}, \ and\ \bibinfo {author} {\bibfnamefont {R.}~\bibnamefont {Zemel}},\ }in\ \href@noop {} {\emph {\bibinfo {booktitle} {International conference on machine learning}}}\ (\bibinfo {organization} {Pmlr},\ \bibinfo {year} {2018})\ pp.\ \bibinfo {pages} {2688--2697}\BibitemShut {NoStop}%
\bibitem [{\citenamefont {Zhang}\ \emph {et~al.}(2019)\citenamefont {Zhang}, \citenamefont {Zhao}, \citenamefont {Liu}, \citenamefont {Wang}, \citenamefont {Tao}, \citenamefont {Xin},\ and\ \citenamefont {Zhang}}]{zhang2019general}%
  \BibitemOpen
  \bibfield  {author} {\bibinfo {author} {\bibfnamefont {Z.}~\bibnamefont {Zhang}}, \bibinfo {author} {\bibfnamefont {Y.}~\bibnamefont {Zhao}}, \bibinfo {author} {\bibfnamefont {J.}~\bibnamefont {Liu}}, \bibinfo {author} {\bibfnamefont {S.}~\bibnamefont {Wang}}, \bibinfo {author} {\bibfnamefont {R.}~\bibnamefont {Tao}}, \bibinfo {author} {\bibfnamefont {R.}~\bibnamefont {Xin}}, \ and\ \bibinfo {author} {\bibfnamefont {J.}~\bibnamefont {Zhang}},\ }\href@noop {} {\bibfield  {journal} {\bibinfo  {journal} {Applied Network Science}\ }\textbf {\bibinfo {volume} {4}},\ \bibinfo {pages} {110} (\bibinfo {year} {2019})}\BibitemShut {NoStop}%
\bibitem [{\citenamefont {Peixoto}(2018)}]{peixoto2018reconstructing}%
  \BibitemOpen
  \bibfield  {author} {\bibinfo {author} {\bibfnamefont {T.~P.}\ \bibnamefont {Peixoto}},\ }\href@noop {} {\bibfield  {journal} {\bibinfo  {journal} {Physical Review X}\ }\textbf {\bibinfo {volume} {8}},\ \bibinfo {pages} {041011} (\bibinfo {year} {2018})}\BibitemShut {NoStop}%
\bibitem [{\citenamefont {Peixoto}(2019)}]{peixoto2019network}%
  \BibitemOpen
  \bibfield  {author} {\bibinfo {author} {\bibfnamefont {T.~P.}\ \bibnamefont {Peixoto}},\ }\href@noop {} {\bibfield  {journal} {\bibinfo  {journal} {Physical review letters}\ }\textbf {\bibinfo {volume} {123}},\ \bibinfo {pages} {128301} (\bibinfo {year} {2019})}\BibitemShut {NoStop}%
\bibitem [{\citenamefont {Casadiego}\ \emph {et~al.}(2017)\citenamefont {Casadiego}, \citenamefont {Nitzan}, \citenamefont {Hallerberg},\ and\ \citenamefont {Timme}}]{casadiego2017model}%
  \BibitemOpen
  \bibfield  {author} {\bibinfo {author} {\bibfnamefont {J.}~\bibnamefont {Casadiego}}, \bibinfo {author} {\bibfnamefont {M.}~\bibnamefont {Nitzan}}, \bibinfo {author} {\bibfnamefont {S.}~\bibnamefont {Hallerberg}}, \ and\ \bibinfo {author} {\bibfnamefont {M.}~\bibnamefont {Timme}},\ }\href@noop {} {\bibfield  {journal} {\bibinfo  {journal} {Nature communications}\ }\textbf {\bibinfo {volume} {8}},\ \bibinfo {pages} {2192} (\bibinfo {year} {2017})}\BibitemShut {NoStop}%
\bibitem [{\citenamefont {Runge}(2018)}]{runge2018causal}%
  \BibitemOpen
  \bibfield  {author} {\bibinfo {author} {\bibfnamefont {J.}~\bibnamefont {Runge}},\ }\href@noop {} {\bibfield  {journal} {\bibinfo  {journal} {Chaos: An Interdisciplinary Journal of Nonlinear Science}\ }\textbf {\bibinfo {volume} {28}} (\bibinfo {year} {2018})}\BibitemShut {NoStop}%
\bibitem [{\citenamefont {Peixoto}(2025)}]{peixoto2025scalable}%
  \BibitemOpen
  \bibfield  {author} {\bibinfo {author} {\bibfnamefont {T.~P.}\ \bibnamefont {Peixoto}},\ }\href@noop {} {\bibfield  {journal} {\bibinfo  {journal} {Proceedings of the Royal Society A: Mathematical, Physical and Engineering Sciences}\ }\textbf {\bibinfo {volume} {481}} (\bibinfo {year} {2025})}\BibitemShut {NoStop}%
\bibitem [{\citenamefont {Wang}\ \emph {et~al.}(2016)\citenamefont {Wang}, \citenamefont {Lai},\ and\ \citenamefont {Grebogi}}]{wang2016data}%
  \BibitemOpen
  \bibfield  {author} {\bibinfo {author} {\bibfnamefont {W.-X.}\ \bibnamefont {Wang}}, \bibinfo {author} {\bibfnamefont {Y.-C.}\ \bibnamefont {Lai}}, \ and\ \bibinfo {author} {\bibfnamefont {C.}~\bibnamefont {Grebogi}},\ }\href@noop {} {\bibfield  {journal} {\bibinfo  {journal} {Physics Reports}\ }\textbf {\bibinfo {volume} {644}},\ \bibinfo {pages} {1} (\bibinfo {year} {2016})}\BibitemShut {NoStop}%
\bibitem [{\citenamefont {Ma}\ \emph {et~al.}(2018)\citenamefont {Ma}, \citenamefont {Chen}, \citenamefont {Lai},\ and\ \citenamefont {Zhang}}]{ma2018statistical}%
  \BibitemOpen
  \bibfield  {author} {\bibinfo {author} {\bibfnamefont {C.}~\bibnamefont {Ma}}, \bibinfo {author} {\bibfnamefont {H.-S.}\ \bibnamefont {Chen}}, \bibinfo {author} {\bibfnamefont {Y.-C.}\ \bibnamefont {Lai}}, \ and\ \bibinfo {author} {\bibfnamefont {H.-F.}\ \bibnamefont {Zhang}},\ }\href@noop {} {\bibfield  {journal} {\bibinfo  {journal} {Physical Review E}\ }\textbf {\bibinfo {volume} {97}},\ \bibinfo {pages} {022301} (\bibinfo {year} {2018})}\BibitemShut {NoStop}%
\bibitem [{\citenamefont {Eagle}\ \emph {et~al.}(2009)\citenamefont {Eagle}, \citenamefont {Pentland},\ and\ \citenamefont {Lazer}}]{eagle2009inferring}%
  \BibitemOpen
  \bibfield  {author} {\bibinfo {author} {\bibfnamefont {N.}~\bibnamefont {Eagle}}, \bibinfo {author} {\bibfnamefont {A.}~\bibnamefont {Pentland}}, \ and\ \bibinfo {author} {\bibfnamefont {D.}~\bibnamefont {Lazer}},\ }\href@noop {} {\bibfield  {journal} {\bibinfo  {journal} {Proceedings of the national academy of sciences}\ }\textbf {\bibinfo {volume} {106}},\ \bibinfo {pages} {15274} (\bibinfo {year} {2009})}\BibitemShut {NoStop}%
\bibitem [{\citenamefont {Moscard{\'o}~Garc{\'\i}a}\ \emph {et~al.}(2025)\citenamefont {Moscard{\'o}~Garc{\'\i}a}, \citenamefont {Aalto}, \citenamefont {Montanari}, \citenamefont {Skupin},\ and\ \citenamefont {Gon{\c{c}}alves}}]{moscardo2025multi}%
  \BibitemOpen
  \bibfield  {author} {\bibinfo {author} {\bibfnamefont {M.}~\bibnamefont {Moscard{\'o}~Garc{\'\i}a}}, \bibinfo {author} {\bibfnamefont {A.}~\bibnamefont {Aalto}}, \bibinfo {author} {\bibfnamefont {A.~N.}\ \bibnamefont {Montanari}}, \bibinfo {author} {\bibfnamefont {A.}~\bibnamefont {Skupin}}, \ and\ \bibinfo {author} {\bibfnamefont {J.}~\bibnamefont {Gon{\c{c}}alves}},\ }\href@noop {} {\bibfield  {journal} {\bibinfo  {journal} {npj Systems Biology and Applications}\ }\textbf {\bibinfo {volume} {11}},\ \bibinfo {pages} {114} (\bibinfo {year} {2025})}\BibitemShut {NoStop}%
\bibitem [{\citenamefont {Zhang}\ \emph {et~al.}(2008)\citenamefont {Zhang}, \citenamefont {Friend}, \citenamefont {Traud}, \citenamefont {Porter}, \citenamefont {Fowler},\ and\ \citenamefont {Mucha}}]{zhang2008community}%
  \BibitemOpen
  \bibfield  {author} {\bibinfo {author} {\bibfnamefont {Y.}~\bibnamefont {Zhang}}, \bibinfo {author} {\bibfnamefont {A.~J.}\ \bibnamefont {Friend}}, \bibinfo {author} {\bibfnamefont {A.~L.}\ \bibnamefont {Traud}}, \bibinfo {author} {\bibfnamefont {M.~A.}\ \bibnamefont {Porter}}, \bibinfo {author} {\bibfnamefont {J.~H.}\ \bibnamefont {Fowler}}, \ and\ \bibinfo {author} {\bibfnamefont {P.~J.}\ \bibnamefont {Mucha}},\ }\href@noop {} {\bibfield  {journal} {\bibinfo  {journal} {Physica A: Statistical Mechanics and its Applications}\ }\textbf {\bibinfo {volume} {387}},\ \bibinfo {pages} {1705} (\bibinfo {year} {2008})}\BibitemShut {NoStop}%
\bibitem [{\citenamefont {Yan}\ \emph {et~al.}(2025)\citenamefont {Yan}, \citenamefont {Yang}, \citenamefont {Wu}, \citenamefont {Liu}, \citenamefont {Zhang}, \citenamefont {Li}, \citenamefont {Tan},\ and\ \citenamefont {Wu}}]{yan2025efficient}%
  \BibitemOpen
  \bibfield  {author} {\bibinfo {author} {\bibfnamefont {Y.}~\bibnamefont {Yan}}, \bibinfo {author} {\bibfnamefont {Q.}~\bibnamefont {Yang}}, \bibinfo {author} {\bibfnamefont {Y.}~\bibnamefont {Wu}}, \bibinfo {author} {\bibfnamefont {H.}~\bibnamefont {Liu}}, \bibinfo {author} {\bibfnamefont {M.}~\bibnamefont {Zhang}}, \bibinfo {author} {\bibfnamefont {H.}~\bibnamefont {Li}}, \bibinfo {author} {\bibfnamefont {K.~C.}\ \bibnamefont {Tan}}, \ and\ \bibinfo {author} {\bibfnamefont {J.}~\bibnamefont {Wu}},\ }\href@noop {} {\bibfield  {journal} {\bibinfo  {journal} {Nature communications}\ }\textbf {\bibinfo {volume} {16}},\ \bibinfo {pages} {8651} (\bibinfo {year} {2025})}\BibitemShut {NoStop}%
\bibitem [{\citenamefont {Pastor-Satorras}\ \emph {et~al.}(2015)\citenamefont {Pastor-Satorras}, \citenamefont {Castellano}, \citenamefont {Van~Mieghem},\ and\ \citenamefont {Vespignani}}]{pastor2015epidemic}%
  \BibitemOpen
  \bibfield  {author} {\bibinfo {author} {\bibfnamefont {R.}~\bibnamefont {Pastor-Satorras}}, \bibinfo {author} {\bibfnamefont {C.}~\bibnamefont {Castellano}}, \bibinfo {author} {\bibfnamefont {P.}~\bibnamefont {Van~Mieghem}}, \ and\ \bibinfo {author} {\bibfnamefont {A.}~\bibnamefont {Vespignani}},\ }\href@noop {} {\bibfield  {journal} {\bibinfo  {journal} {Reviews of modern physics}\ }\textbf {\bibinfo {volume} {87}},\ \bibinfo {pages} {925} (\bibinfo {year} {2015})}\BibitemShut {NoStop}%
\bibitem [{\citenamefont {Bao}\ and\ \citenamefont {Wu}(2025)}]{bao2025epidemic}%
  \BibitemOpen
  \bibfield  {author} {\bibinfo {author} {\bibfnamefont {H.}~\bibnamefont {Bao}}\ and\ \bibinfo {author} {\bibfnamefont {X.}~\bibnamefont {Wu}},\ }\href@noop {} {\bibfield  {journal} {\bibinfo  {journal} {Nonlinear Dynamics}\ }\textbf {\bibinfo {volume} {113}},\ \bibinfo {pages} {5881} (\bibinfo {year} {2025})}\BibitemShut {NoStop}%
\bibitem [{\citenamefont {Shafer}(2020)}]{shafer2020mathematical}%
  \BibitemOpen
  \bibfield  {author} {\bibinfo {author} {\bibfnamefont {G.}~\bibnamefont {Shafer}},\ }\href@noop {} {\emph {\bibinfo {title} {A mathematical theory of evidence}}}\ (\bibinfo  {publisher} {Princeton university press},\ \bibinfo {year} {2020})\BibitemShut {NoStop}%
\bibitem [{\citenamefont {Dempster}(1968)}]{dempster1968upper}%
  \BibitemOpen
  \bibfield  {author} {\bibinfo {author} {\bibfnamefont {A.~P.}\ \bibnamefont {Dempster}},\ }\href@noop {} {\bibfield  {journal} {\bibinfo  {journal} {The Annals of Mathematical Statistics}\ ,\ \bibinfo {pages} {957}} (\bibinfo {year} {1968})}\BibitemShut {NoStop}%
\bibitem [{\citenamefont {Deng}(2015)}]{deng2015generalized}%
  \BibitemOpen
  \bibfield  {author} {\bibinfo {author} {\bibfnamefont {Y.}~\bibnamefont {Deng}},\ }\href@noop {} {\bibfield  {journal} {\bibinfo  {journal} {Applied Intelligence}\ }\textbf {\bibinfo {volume} {43}},\ \bibinfo {pages} {530} (\bibinfo {year} {2015})}\BibitemShut {NoStop}%
\bibitem [{\citenamefont {Barab{\'a}si}\ and\ \citenamefont {Albert}(1999)}]{barabasi1999emergence}%
  \BibitemOpen
  \bibfield  {author} {\bibinfo {author} {\bibfnamefont {A.-L.}\ \bibnamefont {Barab{\'a}si}}\ and\ \bibinfo {author} {\bibfnamefont {R.}~\bibnamefont {Albert}},\ }\href@noop {} {\bibfield  {journal} {\bibinfo  {journal} {science}\ }\textbf {\bibinfo {volume} {286}},\ \bibinfo {pages} {509} (\bibinfo {year} {1999})}\BibitemShut {NoStop}%
\bibitem [{\citenamefont {ERDdS}\ and\ \citenamefont {R\&wi}(1959)}]{erdds1959random}%
  \BibitemOpen
  \bibfield  {author} {\bibinfo {author} {\bibfnamefont {P.}~\bibnamefont {ERDdS}}\ and\ \bibinfo {author} {\bibfnamefont {A.}~\bibnamefont {R\&wi}},\ }\href@noop {} {\bibfield  {journal} {\bibinfo  {journal} {Publ. math. debrecen}\ }\textbf {\bibinfo {volume} {6}},\ \bibinfo {pages} {18} (\bibinfo {year} {1959})}\BibitemShut {NoStop}%
\bibitem [{\citenamefont {Watts}\ and\ \citenamefont {Strogatz}(1998)}]{watts1998collective}%
  \BibitemOpen
  \bibfield  {author} {\bibinfo {author} {\bibfnamefont {D.~J.}\ \bibnamefont {Watts}}\ and\ \bibinfo {author} {\bibfnamefont {S.~H.}\ \bibnamefont {Strogatz}},\ }\href@noop {} {\bibfield  {journal} {\bibinfo  {journal} {nature}\ }\textbf {\bibinfo {volume} {393}},\ \bibinfo {pages} {440} (\bibinfo {year} {1998})}\BibitemShut {NoStop}%
\bibitem [{\citenamefont {Dempster}(1967)}]{Dempster1967Upper}%
  \BibitemOpen
  \bibfield  {author} {\bibinfo {author} {\bibfnamefont {A.~P.}\ \bibnamefont {Dempster}},\ }\href {\doibase 10.1214/aoms/1177698950} {\bibfield  {journal} {\bibinfo  {journal} {The Annals of Mathematical Statistics}\ }\textbf {\bibinfo {volume} {38}},\ \bibinfo {pages} {325 } (\bibinfo {year} {1967})}\BibitemShut {NoStop}%
\bibitem [{\citenamefont {Deng}(2022)}]{deng2022random}%
  \BibitemOpen
  \bibfield  {author} {\bibinfo {author} {\bibfnamefont {Y.}~\bibnamefont {Deng}},\ }\href@noop {} {\bibfield  {journal} {\bibinfo  {journal} {International Journal of Computers Communications \& Control}\ }\textbf {\bibinfo {volume} {17}} (\bibinfo {year} {2022})}\BibitemShut {NoStop}%
\bibitem [{\citenamefont {Deng}\ \emph {et~al.}(8349)\citenamefont {Deng}, \citenamefont {Deng},\ and\ \citenamefont {Yang}}]{deng2024RPSR}%
  \BibitemOpen
  \bibfield  {author} {\bibinfo {author} {\bibfnamefont {J.}~\bibnamefont {Deng}}, \bibinfo {author} {\bibfnamefont {Y.}~\bibnamefont {Deng}}, \ and\ \bibinfo {author} {\bibfnamefont {J.-B.}\ \bibnamefont {Yang}},\ }\href@noop {} {\bibfield  {journal} {\bibinfo  {journal} {IEEE Transactions on Pattern Analysis and Machine Intelligence}\ } (\bibinfo {year} {2024. DOI: 10.1109/TPAMI.2024.3438349})}\BibitemShut {NoStop}%
\bibitem [{\citenamefont {Li}\ \emph {et~al.}(2024)\citenamefont {Li}, \citenamefont {Li},\ and\ \citenamefont {Zhang}}]{li2024new}%
  \BibitemOpen
  \bibfield  {author} {\bibinfo {author} {\bibfnamefont {M.}~\bibnamefont {Li}}, \bibinfo {author} {\bibfnamefont {L.}~\bibnamefont {Li}}, \ and\ \bibinfo {author} {\bibfnamefont {Q.}~\bibnamefont {Zhang}},\ }\href@noop {} {\bibfield  {journal} {\bibinfo  {journal} {Information Sciences}\ }\textbf {\bibinfo {volume} {677}},\ \bibinfo {pages} {120883} (\bibinfo {year} {2024})}\BibitemShut {NoStop}%
\bibitem [{\citenamefont {Zadeh}(1965)}]{zadeh1965fuzzy}%
  \BibitemOpen
  \bibfield  {author} {\bibinfo {author} {\bibfnamefont {L.~A.}\ \bibnamefont {Zadeh}},\ }\href@noop {} {\bibfield  {journal} {\bibinfo  {journal} {Information and control}\ }\textbf {\bibinfo {volume} {8}},\ \bibinfo {pages} {338} (\bibinfo {year} {1965})}\BibitemShut {NoStop}%
\bibitem [{\citenamefont {Liu}\ and\ \citenamefont {Deng}(2020)}]{liu2020determine}%
  \BibitemOpen
  \bibfield  {author} {\bibinfo {author} {\bibfnamefont {F.}~\bibnamefont {Liu}}\ and\ \bibinfo {author} {\bibfnamefont {Y.}~\bibnamefont {Deng}},\ }\href@noop {} {\bibfield  {journal} {\bibinfo  {journal} {IEEE Transactions on Fuzzy Systems}\ }\textbf {\bibinfo {volume} {29}},\ \bibinfo {pages} {986} (\bibinfo {year} {2020})}\BibitemShut {NoStop}%
\bibitem [{\citenamefont {Seiti}\ \emph {et~al.}(2018)\citenamefont {Seiti}, \citenamefont {Hafezalkotob}, \citenamefont {Najafi},\ and\ \citenamefont {Khalaj}}]{seiti2018risk}%
  \BibitemOpen
  \bibfield  {author} {\bibinfo {author} {\bibfnamefont {H.}~\bibnamefont {Seiti}}, \bibinfo {author} {\bibfnamefont {A.}~\bibnamefont {Hafezalkotob}}, \bibinfo {author} {\bibfnamefont {S.~E.}\ \bibnamefont {Najafi}}, \ and\ \bibinfo {author} {\bibfnamefont {M.}~\bibnamefont {Khalaj}},\ }\href@noop {} {\bibfield  {journal} {\bibinfo  {journal} {Journal of Intelligent \& Fuzzy Systems}\ }\textbf {\bibinfo {volume} {35}},\ \bibinfo {pages} {1419} (\bibinfo {year} {2018})}\BibitemShut {NoStop}%
\bibitem [{\citenamefont {Liu}\ \emph {et~al.}(2019)\citenamefont {Liu}, \citenamefont {Liu}, \citenamefont {Dezert},\ and\ \citenamefont {Cuzzolin}}]{liu2019evidence}%
  \BibitemOpen
  \bibfield  {author} {\bibinfo {author} {\bibfnamefont {Z.-G.}\ \bibnamefont {Liu}}, \bibinfo {author} {\bibfnamefont {Y.}~\bibnamefont {Liu}}, \bibinfo {author} {\bibfnamefont {J.}~\bibnamefont {Dezert}}, \ and\ \bibinfo {author} {\bibfnamefont {F.}~\bibnamefont {Cuzzolin}},\ }\href@noop {} {\bibfield  {journal} {\bibinfo  {journal} {IEEE Transactions on Fuzzy Systems}\ }\textbf {\bibinfo {volume} {28}},\ \bibinfo {pages} {618} (\bibinfo {year} {2019})}\BibitemShut {NoStop}%
\bibitem [{\citenamefont {Szab{\'o}}\ and\ \citenamefont {Fath}(2007)}]{szabo2007evolutionary}%
  \BibitemOpen
  \bibfield  {author} {\bibinfo {author} {\bibfnamefont {G.}~\bibnamefont {Szab{\'o}}}\ and\ \bibinfo {author} {\bibfnamefont {G.}~\bibnamefont {Fath}},\ }\href@noop {} {\bibfield  {journal} {\bibinfo  {journal} {Physics reports}\ }\textbf {\bibinfo {volume} {446}},\ \bibinfo {pages} {97} (\bibinfo {year} {2007})}\BibitemShut {NoStop}%
\bibitem [{\citenamefont {Granovetter}(1978)}]{granovetter1978threshold}%
  \BibitemOpen
  \bibfield  {author} {\bibinfo {author} {\bibfnamefont {M.}~\bibnamefont {Granovetter}},\ }\href@noop {} {\bibfield  {journal} {\bibinfo  {journal} {American journal of sociology}\ }\textbf {\bibinfo {volume} {83}},\ \bibinfo {pages} {1420} (\bibinfo {year} {1978})}\BibitemShut {NoStop}%
\bibitem [{\citenamefont {Kermack}\ and\ \citenamefont {McKendrick}(1927)}]{kermack1927contribution}%
  \BibitemOpen
  \bibfield  {author} {\bibinfo {author} {\bibfnamefont {W.~O.}\ \bibnamefont {Kermack}}\ and\ \bibinfo {author} {\bibfnamefont {A.~G.}\ \bibnamefont {McKendrick}},\ }\href@noop {} {\bibfield  {journal} {\bibinfo  {journal} {Proceedings of the royal society of london. Series A, Containing papers of a mathematical and physical character}\ }\textbf {\bibinfo {volume} {115}},\ \bibinfo {pages} {700} (\bibinfo {year} {1927})}\BibitemShut {NoStop}%
\bibitem [{\citenamefont {Harris}(1974)}]{10.1214/aop/1176996493}%
  \BibitemOpen
  \bibfield  {author} {\bibinfo {author} {\bibfnamefont {T.~E.}\ \bibnamefont {Harris}},\ }\href {\doibase 10.1214/aop/1176996493} {\bibfield  {journal} {\bibinfo  {journal} {The Annals of Probability}\ }\textbf {\bibinfo {volume} {2}},\ \bibinfo {pages} {969 } (\bibinfo {year} {1974})}\BibitemShut {NoStop}%
\bibitem [{\citenamefont {Zhan}\ \emph {et~al.}(2025)\citenamefont {Zhan}, \citenamefont {Mei}, \citenamefont {Xie}, \citenamefont {Lv}, \citenamefont {Liu},\ and\ \citenamefont {Zhang}}]{zhan2025modeling}%
  \BibitemOpen
  \bibfield  {author} {\bibinfo {author} {\bibfnamefont {X.-X.}\ \bibnamefont {Zhan}}, \bibinfo {author} {\bibfnamefont {G.}~\bibnamefont {Mei}}, \bibinfo {author} {\bibfnamefont {C.}~\bibnamefont {Xie}}, \bibinfo {author} {\bibfnamefont {F.}~\bibnamefont {Lv}}, \bibinfo {author} {\bibfnamefont {C.}~\bibnamefont {Liu}}, \ and\ \bibinfo {author} {\bibfnamefont {Z.-K.}\ \bibnamefont {Zhang}},\ }\href@noop {} {\bibfield  {journal} {\bibinfo  {journal} {Chaos, Solitons \& Fractals}\ }\textbf {\bibinfo {volume} {199}},\ \bibinfo {pages} {116672} (\bibinfo {year} {2025})}\BibitemShut {NoStop}%
\bibitem [{\citenamefont {Chen}\ \emph {et~al.}(2024)\citenamefont {Chen}, \citenamefont {Xi}, \citenamefont {Dong}, \citenamefont {Zhao}, \citenamefont {Li}, \citenamefont {Liu},\ and\ \citenamefont {Cui}}]{chen2024identifying}%
  \BibitemOpen
  \bibfield  {author} {\bibinfo {author} {\bibfnamefont {L.}~\bibnamefont {Chen}}, \bibinfo {author} {\bibfnamefont {Y.}~\bibnamefont {Xi}}, \bibinfo {author} {\bibfnamefont {L.}~\bibnamefont {Dong}}, \bibinfo {author} {\bibfnamefont {M.}~\bibnamefont {Zhao}}, \bibinfo {author} {\bibfnamefont {C.}~\bibnamefont {Li}}, \bibinfo {author} {\bibfnamefont {X.}~\bibnamefont {Liu}}, \ and\ \bibinfo {author} {\bibfnamefont {X.}~\bibnamefont {Cui}},\ }\href@noop {} {\bibfield  {journal} {\bibinfo  {journal} {Information Processing \& Management}\ }\textbf {\bibinfo {volume} {61}},\ \bibinfo {pages} {103775} (\bibinfo {year} {2024})}\BibitemShut {NoStop}%
\bibitem [{\citenamefont {Yin}\ \emph {et~al.}(2024)\citenamefont {Yin}, \citenamefont {Li}, \citenamefont {Wang}, \citenamefont {Lang}, \citenamefont {Hao},\ and\ \citenamefont {Zhang}}]{yin2024identifying}%
  \BibitemOpen
  \bibfield  {author} {\bibinfo {author} {\bibfnamefont {R.}~\bibnamefont {Yin}}, \bibinfo {author} {\bibfnamefont {L.}~\bibnamefont {Li}}, \bibinfo {author} {\bibfnamefont {Y.}~\bibnamefont {Wang}}, \bibinfo {author} {\bibfnamefont {C.}~\bibnamefont {Lang}}, \bibinfo {author} {\bibfnamefont {Z.}~\bibnamefont {Hao}}, \ and\ \bibinfo {author} {\bibfnamefont {L.}~\bibnamefont {Zhang}},\ }\href@noop {} {\bibfield  {journal} {\bibinfo  {journal} {Chaos, Solitons \& Fractals}\ }\textbf {\bibinfo {volume} {180}},\ \bibinfo {pages} {114487} (\bibinfo {year} {2024})}\BibitemShut {NoStop}%
\bibitem [{\citenamefont {Moore}\ and\ \citenamefont {Newman}(2000)}]{PhysRevE.61.5678}%
  \BibitemOpen
  \bibfield  {author} {\bibinfo {author} {\bibfnamefont {C.}~\bibnamefont {Moore}}\ and\ \bibinfo {author} {\bibfnamefont {M.~E.~J.}\ \bibnamefont {Newman}},\ }\href {\doibase 10.1103/PhysRevE.61.5678} {\bibfield  {journal} {\bibinfo  {journal} {Phys. Rev. E}\ }\textbf {\bibinfo {volume} {61}},\ \bibinfo {pages} {5678} (\bibinfo {year} {2000})}\BibitemShut {NoStop}%
\bibitem [{\citenamefont {Awolude}\ \emph {et~al.}(2025)\citenamefont {Awolude}, \citenamefont {Don},\ and\ \citenamefont {Cator}}]{PhysRevE.111.024315}%
  \BibitemOpen
  \bibfield  {author} {\bibinfo {author} {\bibfnamefont {O.~S.}\ \bibnamefont {Awolude}}, \bibinfo {author} {\bibfnamefont {H.}~\bibnamefont {Don}}, \ and\ \bibinfo {author} {\bibfnamefont {E.}~\bibnamefont {Cator}},\ }\href {\doibase 10.1103/PhysRevE.111.024315} {\bibfield  {journal} {\bibinfo  {journal} {Phys. Rev. E}\ }\textbf {\bibinfo {volume} {111}},\ \bibinfo {pages} {024315} (\bibinfo {year} {2025})}\BibitemShut {NoStop}%
\bibitem [{\citenamefont {Li}\ \emph {et~al.}(2018)\citenamefont {Li}, \citenamefont {Zhang},\ and\ \citenamefont {Deng}}]{li2018evidential}%
  \BibitemOpen
  \bibfield  {author} {\bibinfo {author} {\bibfnamefont {M.}~\bibnamefont {Li}}, \bibinfo {author} {\bibfnamefont {Q.}~\bibnamefont {Zhang}}, \ and\ \bibinfo {author} {\bibfnamefont {Y.}~\bibnamefont {Deng}},\ }\href@noop {} {\bibfield  {journal} {\bibinfo  {journal} {Chaos, Solitons \& Fractals}\ }\textbf {\bibinfo {volume} {117}},\ \bibinfo {pages} {283} (\bibinfo {year} {2018})}\BibitemShut {NoStop}%
\bibitem [{\citenamefont {Zachary}(1977)}]{zachary1977information}%
  \BibitemOpen
  \bibfield  {author} {\bibinfo {author} {\bibfnamefont {W.~W.}\ \bibnamefont {Zachary}},\ }\href@noop {} {\bibfield  {journal} {\bibinfo  {journal} {Journal of anthropological research}\ }\textbf {\bibinfo {volume} {33}},\ \bibinfo {pages} {452} (\bibinfo {year} {1977})}\BibitemShut {NoStop}%
\bibitem [{\citenamefont {Zhang}\ and\ \citenamefont {Li}(2022)}]{zhang2022betweenness}%
  \BibitemOpen
  \bibfield  {author} {\bibinfo {author} {\bibfnamefont {Q.}~\bibnamefont {Zhang}}\ and\ \bibinfo {author} {\bibfnamefont {M.}~\bibnamefont {Li}},\ }\href@noop {} {\bibfield  {journal} {\bibinfo  {journal} {Chaos, Solitons \& Fractals}\ }\textbf {\bibinfo {volume} {161}},\ \bibinfo {pages} {112264} (\bibinfo {year} {2022})}\BibitemShut {NoStop}%
\bibitem [{\citenamefont {Xian}\ \emph {et~al.}(2025{\natexlab{b}})\citenamefont {Xian}, \citenamefont {Li},\ and\ \citenamefont {Zhang}}]{xian2025k}%
  \BibitemOpen
  \bibfield  {author} {\bibinfo {author} {\bibfnamefont {Y.}~\bibnamefont {Xian}}, \bibinfo {author} {\bibfnamefont {M.}~\bibnamefont {Li}}, \ and\ \bibinfo {author} {\bibfnamefont {Q.}~\bibnamefont {Zhang}},\ }\href@noop {} {\bibfield  {journal} {\bibinfo  {journal} {Physica A: Statistical Mechanics and its Applications}\ ,\ \bibinfo {pages} {130859}} (\bibinfo {year} {2025}{\natexlab{b}})}\BibitemShut {NoStop}%
\bibitem [{\citenamefont {Rossi}\ and\ \citenamefont {Ahmed}(2015)}]{nr}%
  \BibitemOpen
  \bibfield  {author} {\bibinfo {author} {\bibfnamefont {R.~A.}\ \bibnamefont {Rossi}}\ and\ \bibinfo {author} {\bibfnamefont {N.~K.}\ \bibnamefont {Ahmed}},\ }in\ \href {http://networkrepository.com} {\emph {\bibinfo {booktitle} {Proceedings of the Twenty-Ninth AAAI Conference on Artificial Intelligence}}}\ (\bibinfo {year} {2015})\BibitemShut {NoStop}%
\bibitem [{\citenamefont {Cho}\ \emph {et~al.}(2014)\citenamefont {Cho}, \citenamefont {Shin}, \citenamefont {Hwang}, \citenamefont {Kim}, \citenamefont {Shim}, \citenamefont {Kim}, \citenamefont {Kim},\ and\ \citenamefont {Lee}}]{cho2014wormnet}%
  \BibitemOpen
  \bibfield  {author} {\bibinfo {author} {\bibfnamefont {A.}~\bibnamefont {Cho}}, \bibinfo {author} {\bibfnamefont {J.}~\bibnamefont {Shin}}, \bibinfo {author} {\bibfnamefont {S.}~\bibnamefont {Hwang}}, \bibinfo {author} {\bibfnamefont {C.}~\bibnamefont {Kim}}, \bibinfo {author} {\bibfnamefont {H.}~\bibnamefont {Shim}}, \bibinfo {author} {\bibfnamefont {H.}~\bibnamefont {Kim}}, \bibinfo {author} {\bibfnamefont {H.}~\bibnamefont {Kim}}, \ and\ \bibinfo {author} {\bibfnamefont {I.}~\bibnamefont {Lee}},\ }\href@noop {} {\bibfield  {journal} {\bibinfo  {journal} {Nucleic acids research}\ }\textbf {\bibinfo {volume} {42}},\ \bibinfo {pages} {W76} (\bibinfo {year} {2014})}\BibitemShut {NoStop}%
\bibitem [{\citenamefont {Ahmed}\ \emph {et~al.}(2010)\citenamefont {Ahmed}, \citenamefont {Berchmans}, \citenamefont {Neville},\ and\ \citenamefont {Kompella}}]{ahmed2010time}%
  \BibitemOpen
  \bibfield  {author} {\bibinfo {author} {\bibfnamefont {N.}~\bibnamefont {Ahmed}}, \bibinfo {author} {\bibfnamefont {F.}~\bibnamefont {Berchmans}}, \bibinfo {author} {\bibfnamefont {J.}~\bibnamefont {Neville}}, \ and\ \bibinfo {author} {\bibfnamefont {R.}~\bibnamefont {Kompella}},\ }in\ \href@noop {} {\emph {\bibinfo {booktitle} {SIGKDD MLG}}}\ (\bibinfo {year} {2010})\ pp.\ \bibinfo {pages} {1--9}\BibitemShut {NoStop}%
\bibitem [{\citenamefont {Rossi}\ \emph {et~al.}(2012)\citenamefont {Rossi}, \citenamefont {Gleich}, \citenamefont {Gebremedhin},\ and\ \citenamefont {Patwary}}]{rossi2012fastclique}%
  \BibitemOpen
  \bibfield  {author} {\bibinfo {author} {\bibfnamefont {R.~A.}\ \bibnamefont {Rossi}}, \bibinfo {author} {\bibfnamefont {D.~F.}\ \bibnamefont {Gleich}}, \bibinfo {author} {\bibfnamefont {A.~H.}\ \bibnamefont {Gebremedhin}}, \ and\ \bibinfo {author} {\bibfnamefont {M.~A.}\ \bibnamefont {Patwary}},\ }\href@noop {} {\bibfield  {journal} {\bibinfo  {journal} {arXiv preprint arXiv:1210.5802}\ ,\ \bibinfo {pages} {1}} (\bibinfo {year} {2012})}\BibitemShut {NoStop}%
\bibitem [{\citenamefont {Rossi}\ \emph {et~al.}(2014)\citenamefont {Rossi}, \citenamefont {Gleich}, \citenamefont {Gebremedhin},\ and\ \citenamefont {Patwary}}]{rossi2014pmc-www}%
  \BibitemOpen
  \bibfield  {author} {\bibinfo {author} {\bibfnamefont {R.~A.}\ \bibnamefont {Rossi}}, \bibinfo {author} {\bibfnamefont {D.~F.}\ \bibnamefont {Gleich}}, \bibinfo {author} {\bibfnamefont {A.~H.}\ \bibnamefont {Gebremedhin}}, \ and\ \bibinfo {author} {\bibfnamefont {M.~A.}\ \bibnamefont {Patwary}},\ }in\ \href@noop {} {\emph {\bibinfo {booktitle} {Proceedings of the 23rd International Conference on World Wide Web (WWW)}}}\ (\bibinfo {year} {2014})\BibitemShut {NoStop}%
\bibitem [{\citenamefont {{Truthy}}()}]{truthy}%
  \BibitemOpen
  \bibfield  {author} {\bibinfo {author} {\bibnamefont {{Truthy}}},\ }\href@noop {} {\enquote {\bibinfo {title} {Information diffusion research at indiana university},}\ }\bibinfo {note} {{\url{http://truthy.indiana.edu/}}. Accessed 10/20/12.}\BibitemShut {Stop}%
\end{thebibliography}%
\end{document}